\documentclass[useAMS,usenatbib]{mn2e}
\usepackage[ansinew]{inputenc}
\usepackage{graphicx}
\usepackage{rotating}
\usepackage{tabto}
\usepackage[flushleft]{threeparttable}
\usepackage{pdflscape}
\usepackage{float}
\usepackage{xcolor}
\usepackage{caption}
\usepackage{color}
\usepackage{url}
\usepackage{aas_macros}
\usepackage{hyperref}

\title[Escape and evolution of Titan's atmosphere]
  {Escape and evolution of Titan's N$_2$ atmosphere constrained by $^{14}$N/$^{15}$N isotope ratios}
\author[N. V. Erkaev  et al.]
{N.V.~Erkaev$^{1,2}$, M.~Scherf$^3$, S.E.~Thaller$^4$ H.~Lammer$^3$, A.V.~Mezentsev$^{2}$,
\newauthor V.A., Ivanov$^{2}$, and K.E.~Mandt$^{5}$
 \\
 \\
$^1$Institute of Computational Modelling SB RAS, 660036, Krasnoyarsk, Russian Federation\\
$^2$Siberian Federal University, Krasnoyarsk, Russian Federation\\
$^3$Space Research Institute, Austrian Academy of Sciences, Schmiedlstr. 6, A-8042, Graz, Austria\\
$^4$Institute of Physics/IGAM, University of Graz, Austria\\
$^5$John Hopkins University, Baltimore, Maryland 21218, USA}
\date{Released 2020}

\pagerange{\pageref{firstpage}--\pageref{lastpage}} \pubyear{2019}

\def\LaTeX{L\kern-.36em\raise.3ex\hbox{a}\kern-.15em
    T\kern-.1667em\lower.7ex\hbox{E}\kern-.125emX}

\begin{document}

\label{firstpage}

\maketitle

\begin{abstract}
We apply a 1D upper atmosphere model to study thermal escape of nitrogen over Titan's history. Significant thermal escape should have occurred very early for solar EUV fluxes 100 to 400 times higher than today with escape rates as high as $\approx 1.5\times 10^{28}$\,s$^{-1}$ and $\approx 4.5\times 10^{29}$\,s$^{-1}$, respectively, while today it is $\approx 7.5\times 10^{17}$\,s$^{-1}$. Depending on whether the Sun originated as a slow, moderate or fast rotator, thermal escape was the dominant escape process for the first 100 to 1000 Myr after the formation of the solar system. If Titan's atmosphere originated that early, it could have lost between $\approx\,0.5\,-\,16$ times its present atmospheric mass depending on the Sun's rotational evolution. We also investigated the mass-balance parameter space for an outgassing of Titan's nitrogen through decomposition of NH$_3$-ices in its deep interior. Our study indicates that, if Titan's atmosphere originated at the beginning, it could have only survived until today if the Sun was a slow rotator. In other cases, the escape would have been too strong for the degassed nitrogen to survive until present-day, implying later outgassing or an additional nitrogen source. An endogenic origin of Titan's nitrogen partially through NH$_3$-ices is consistent with its initial fractionation of $^{14}$N/$^{15}$N\,$\approx$\,166\,--\,172, or lower if photochemical removal was relevant for longer than the last $\approx$\,1,000\,Myr. Since this ratio is slightly above the ratio of cometary ammonia, some of Titan's nitrogen might have originated from refractory organics.
\end{abstract}

\begin{keywords}
planets and satellites: atmospheres -- planets and satellites: physical
evolution -- ultraviolet: planetary systems -- stars: ultraviolet -- hydrodynamics
\end{keywords}

\section{INTRODUCTION}

Titan the largest satellite of Saturn developed an N$_2$ atmosphere that is thicker than that of the Earth. In Earth's case its nitrogen atmosphere originated from chondritic bodies \citep{Marty2012,Fueri2015,Harries2015}, while the source and evolution of N$_2$ atmospheres of icy bodies like Titan might be different. The isotopic ratio of $^{14}$N/$^{15}$N keeps information on the origin of the building blocks and also on the evolution of the nitrogen atmosphere. There are several primordial nitrogen reservoirs in the solar system with different initial $^{14}$N/$^{15}$N ratios. Earth's $^{14}$N/$^{15}$N ratio is 272, while Titan's atmospheric nitrogen isotope ratio is $167.7\pm0.6$ \citep[e.g.][]{Niemann2010,Mandt2009,Mandt2014,Fueri2015}. Previous studies predicted that non-thermal and thermal atmospheric escape \citep{Lammer2001,Lammer2008,Penz2005,Lunine1999} of the lighter $^{14}$N isotope during the satellites lifetime modified the isotope ratio from an Earth-like value to its observed ratio today. Depending on assumed homopause-exobase distances, the EUV flux -- the wavelenght range between $\approx$\,10\,--120\,nm -- and solar wind conditions of the young Sun, these studies predicted that Titan's early atmosphere may have lost and could have been more than 30 times denser than at present-day. Besides atmospheric escape the isotopes are also fractionated over geologic time scales by diffusion, photochemistry and photochemical processes \citep{Krasnopolsky2016,Liang2007}. Diffusion and atmospheric escape processes preferentially remove the lighter isotopes, while photolysis of nitrogen preferentially removes the heavier isotopes.

\citet{Mandt2009} investigated the isotopic evolution of $^{14}$N/$^{15}$N, $^{12}$C/$^{13}$C and D/H isotopes in Titan's atmosphere and concluded that,
if one considers early EUV-driven hydrodynamic escape caused by the higher activity of the young Sun, the initial $^{14}$N/$^{15}$N ratio in N$_2$ most likely did not  change much from its current value as result of atmospheric processes since the time for which hydrodynamic escape may have been active did not last long enough to fractionate N$_2$. Moreover, these authors also concluded that, even though low-rate loss processes such as Jeans escape, photochemical removal or atmospheric sputtering are efficient fractionators, they take a very long time to influence such a large nitrogen inventory.

Due to the difficulty in fractionating Titan's nitrogen isotopes from the chondritic value in Earth's atmosphere to the much lower observed ratio in Titan's present-day atmosphere, \citet{Mandt2014} suggested that the upper limit for the primordial ratio of $^{14}$N/$^{15}$N for Titan is limited to $\le$190, which also fits approximately with measurements of cometary NH$_3$ and NH$_2$ species \citep[e.g.][]{Rousselot2014,Shinnaka2014,Shinnaka2016}. Therefore, \citet{Mandt2014} concluded that Titan's N$_2$ originated from ammonia ices in the cool protosolar nebula and not from building blocks that formed within the Saturnian subnebula.

One should note that the above mentioned studies did not apply a sub-/transonic upper atmosphere model that calculates the atmosphere structures, temperature profiles, and homopause-exobase distances as a function of the expected EUV flux evolution of the young Sun \citep{Tu2015,Lammer2018}. Due to this, their conclusions related to atmospheric escape are only based on rough assumptions.

Recent advances in the research on the evolution of the solar EUV flux \citep{Tu2015} and mass loss \citep{Johnstone2015I,johnstone2015II} provide potential evolutionary tracks that can be incorporated into evolutionary models of atmospheric escape at Titan. As \citet{Tu2015} illustrate, the evolution of the EUV flux of a star can follow different evolutionary tracks that are correlated with the evolution of a stars rotation rate. If the initial rotation rate is high, the EUV flux will stay in its saturation phase for a longer time than for a moderate or a slow rotator, and so does the star's mass loss. These winds also lead to a slowing down of the rotation rate due to the removal of angular momentum \citep{Weber1967,Kraft1967}, leading to the convergence of the different rotational tracks after about a billion years for G-type stars \citep{Johnstone2015I}. By now it is not entirely clear whether our Sun was a slowly, moderately, or fast rotating young G-type star or something in-between. There are, however, some recent studies indicating that it cannot have been a fast, but a slow, or a slow to moderate rotator \citep{Saxena2019,Lammer2020,Johnstone2020}. \citet{Saxena2019} reconstructed the loss of K and Na from the Lunar surface due to sputtering from Coronal Mass Ejections (CMEs) for different rotational tracks of the young Sun. They found that a slow rotator can reproduce the Lunar surface composition, while for the other rotators the depletion in volatiles would be too high. This is in agreement with a study by \citet{Lammer2020}, who investigated the fractionation of Ar and Ne isotopes in the atmospheres of the Earth and Venus due to the escape of a primordial hydrogen atmosphere that was accreted from the solar nebula, and found that only a slow, or slow to moderate rotator can reproduce the present-day \textbf{ratios}. Furthermore, \citet{Johnstone2020} found, through modelling of the upper atmosphere of the Earth for different CO$_2$/N$_2$ mixing ratios and EUV fluxes, that the terrestrial nitrogen-dominated atmosphere would not have been stable if the Sun weren't a slow rotator. Taken these results together indicates that the Sun was a weakly active young G-type star which also has indications for the evolution of Titan's atmosphere, as we will illustrate in the upcoming sections.

The aim of this work is to infer the most realistic atmospheric evolution through Titan's history and its initial $^{14}$N/$^{15}$N isotopic ratios based on detailed sub-/transonic upper atmosphere modeling and, in addition, on an estimate of atmospheric sputtering and photochemical processes through time. Here we apply for the first time an upper atmosphere model that studies Jeans and hydrodynamic escape of Titan's nitrogen atmosphere that we exposed to different solar EUV evolution tracks based on slowly, moderately and fast rotating young solar-like G-stars \citep{Tu2015}. We then compare the obtained thermal escape with non-thermal escape processes such as ion-pickup and atmospheric sputtering, but also with photochemical sequestration of N$_2$ into HCN. In Sect.~2 we describe the time-dependent upper atmosphere model approach that covers the sub- and transonic regime, the obtained atmospheric structures, and temperature profiles as a function of solar EUV flux. In Sect.~3 we present and discuss the thermal and non-thermal escape rates and the total nitrogen mass loss through time. This leads to a discussion on the potential building-blocks of Titan's atmosphere and an estimation of the $^{14}$N/$^{15}$N ratio over Titan's history in Sect.~4. The outgassing of nitrogen from the \textbf{satellite's} interior in relation to atmospheric escape and the build up of Titan's atmosphere is discussed in Sect.~4, whereas Sect.~5 concludes the study.

\section{MODELING APPROACH}\label{Sec2}

For the simulation of Titan's upper nitrogen atmosphere evolution, we apply a time-dependent 1-D sub-/transonic model
\citep{Erkaev2015,Erkaev2016,Erkaev2017,Kubyshkina2018} that solves the system of fluid equations for mass, momentum, and energy conservation:
\begin{equation}
\frac{\mathrm{\partial} \rho}{\partial t} + \frac{\partial({\rho V r^2})}{r^2\partial r} = 0,
\end{equation}

\begin{eqnarray}
\frac{\partial (\rho V) }{\partial t} + \frac{\partial \left(r^2 \rho V^2\right)}{r^2\partial r} =
-\rho \nabla U  - \nabla{P},
\end{eqnarray}

\begin{eqnarray}
\frac{\partial \left(\frac{1}{2}\rho V^2+ E +\rho U \right)}{\partial t}
+\frac{\partial V r^2\left(\frac{1}{2}\rho V^2 + E + P + \rho U \right) }{r^2\partial r} =\nonumber\\
Q_{\rm EUV} +  \frac{\partial }{r^2\partial r}\left(r^2 \chi \frac{\partial T}{\partial r}\right).
\end{eqnarray}
Here $\rho$, $V$, $P$, $E$, $T$ are the total mass density, bulk velocity, pressure, thermal energy,
and temperature, respectively;
$\chi$ is the thermal conductivity \citep{Watson1981},
 \begin{equation}
 \chi = 4.45\cdot 10^4 \left(\frac{T}{1000}\right)^{0.7},
 \end{equation}
$U$ is the gravitational potential,
\begin{eqnarray}
U = -\frac{G M_{\rm pl}}{r} ,
\end{eqnarray}
where $G$ is the gravitational constant, $M_{\rm pl}$ is the mass of the planet or satellite,
$Q_{\rm EUV}$ is the stellar EUV volume heating rate, which depends on the stellar EUV flux at
the orbital distance of the Saturn-Titan system at 10\,AU and on the atmospheric density
\begin{equation}
Q_{\rm EUV} = \eta \sigma_{\rm EUV}n\phi_{\rm EUV} ,
\end{equation}
where $n$ is the density of neutral particles, $\sigma_{\rm EUV}$ is the cross section of the EUV absorption, $\eta$ is the heating efficiency, and
$\phi_{\rm EUV}$ is the function describing the EUV flux absorption in the atmosphere,
\begin{equation}
\phi_{\rm EUV}=\frac{1}{4\pi}\int_0^{\pi/2+\arccos(1/r)} J(r,\theta)
2\pi\sin(\theta)d\theta.
\end{equation}
Here, $J(r,\theta)$  is the function of spherical coordinates that describes the spatial variation of the  EUV flux  due to the atmospheric absorption \citep{Erkaev2015}, and $r$ is the radial distance from the planetary
center normalized to the planetary radius $R_{\rm 0}$, i.e.
\begin{eqnarray}
J(r,\theta) = J_\infty \exp(-\tau(r,\theta)), \\
\tau(r,\theta) = \int_{r}^{\infty}{\sigma_{\rm EUV} n(r')\frac{r'}{\sqrt{r'^2 -r^2\sin(\theta)^2}}}d r'.
\end{eqnarray}
$J_\infty$ is the intensity of the incoming EUV radiation far away from the planet.
The heating efficiency $\eta$ is the ratio of the net local heating rate to the rate of the stellar radiative absorption.
Because the current model does not calculate $\eta$ self-consistently as a function of altitude, we assume an average $\eta$ value for the
EUV absorption by N$_2$ of 25\,\% in the collisional regime \citep{Stevens1992,Krasnopolsky1993,Krasnopolsky1999,Tian2005}.
Similar to \citet{Erkaev2013}, \citet{Lammer2013}, and \citet{Lammer2014}, we assume a single wavelength for all photons ($h\nu = 20$\,eV) and use an average
EUV photoabsorption cross sections $\sigma_{\rm EUV}$ for nitrogen molecules of about $3.5\times10^{-17}$ cm$^{2}$  \citep{Galand1999}.

In the hydrodynamic equations the total mass density ($\rho$),  pressure ($P$),
and thermal energy ($E$) are related to the densities of atomic and molecular neutral nitrogen $n_{\rm N}$ and $n_{\rm N_2}$,
ions $n_{\rm N_{}^+}$ and $n_{\rm N_2^+}$, and electrons as follows:
\begin{eqnarray}
\rho=m_{\rm N}\left(n_{\rm N} + n_{\rm N^+}\right) + m_{\rm N_2}\left( n_{\rm N_2} + n_{\rm N_2^+} \right), \\
P=\left(n_{\rm N}+n_{\rm N^+}+n_{\rm N_2}+n_{N_2^+}+n_{\rm e}\right)k_B T, \\
E = \left[ \frac{3}{2} (n_{\rm N} + n_{\rm N^+} + n_e) + \frac{5}{2}( n_{\rm N_2} + n_{\rm N_2^+}) \right] k_B T .
\end{eqnarray}
Here $k_{\rm B}$ is the Boltzmann constant, and $m_{\rm N}$, $m_{\rm N_2}$ are the
masses of the nitrogen atoms and molecules, respectively.

The continuity equations for the number densities of particles
can be written as
\begin{eqnarray}
\frac{\partial \left(n_{\rm N }\right)}{\partial t} + \frac{1}{r^2}\frac{\partial \left(n_{\rm N} v r^2\right)}{\partial r}= S_{\rm N},
\end{eqnarray}
\begin{eqnarray}
\frac{\partial \left(n_{\rm N^+}\right)}{\partial t} + \frac{1}{r^2}\frac{\partial \left(n_{\rm N^+}v r^2\right)}{\partial r}= S_{\rm N^+},
\end{eqnarray}
\begin{eqnarray}
\frac{\partial \left(n_{\rm N_2^+}\right)}{\partial t} + \frac{1}{r^2}\frac{\partial \left(n_{\rm N_2^+}v r^2\right)}{\partial r}= S_{\rm N_2^+},
\end{eqnarray}
where
\begin{eqnarray}
S_{\rm N} = -\nu_{\rm N} n_{\rm N}
+ 2\alpha_{\rm N} n_{e}n_{\rm N^+} + \nonumber \\
\alpha_{\rm N_2}n_{\rm e}n_{\rm N_2^+}
- 2\gamma_{\rm N} n_{tot} n_{\rm N}^2, \nonumber \\
S_{\rm N^+} = \nu_{\rm N} n_{\rm N}
- \alpha_{\rm N} n_{e}n_{\rm N^+}, \nonumber \\
S_{\rm N_2^+} = \nu_{\rm N} n_{\rm N_2} - \alpha_{\rm N_2} n_{e}n_{\rm N_2^+} .
\end{eqnarray}
The electron density is determined by the quasi-neutrality condition
\begin{eqnarray}
n_{\rm e}=n_{\rm N^+}+n_{\rm N_2^+},
\end{eqnarray}
and the total nitrogen density is the sum of partial densities,
\begin{eqnarray}
n_{tot}=n_{\rm N} + n_{\rm N^+} + n_{\rm N_2} + n_{\rm N_2^+}.
\end{eqnarray}

$\alpha_{\rm N}$ is the recombination rate related to the reaction \mbox{N$^{+}+e\rightarrow N$} of $3.6\times 10^{-12} (250/T)^{0.7}$
cm$^{3}$ s$^{-1}$ \citep{Johnstone2018}, $\alpha_{\rm N_2}$ is the dissociation rate of \mbox{N$_2^+ $+$e\rightarrow$N + N}: $\alpha_{\rm N_2}$=$1.98\times 10^{-7} (300/T)^{0.4}$
cm$^{3}$ s$^{-1}$,
$\gamma_{\rm N}$ is the rate of reaction \mbox{N + N $\rightarrow$ N$_2$}: $\gamma_{\rm N}$ = 9.59 $\cdot $ 10$^{-34}$ (480/T)$^{}$.
$\nu_{\rm N}$ is the
 nitrogen ionization rate \citep{Huebner2015}
\begin{eqnarray}
\nu_{\rm N} = 4,77 \cdot 10^{-7} \phi_{\rm EUV} \rm s^{-1}.
\end{eqnarray}

We apply a two-step MacCormack numerical scheme to integrate the system of equations on time which is described in detail in the Appendix.

A stationary radial distribution of the atmospheric quantities can be obtained
as a result of time relaxation.

The hydrodynamic equations are not applicable in the region, where the ratio of the mean free path of particles between
collisions $\ell_c = 1/(\sigma_c n) $ to the
length scale of the pressure variation $d = (d \ln(P)/ dr)^{-1}$ exceeds 1.
This ratio is called Knudsen number $Kn = 1/(d \, \sigma_c n))$,
where $\sigma_c$ is the collision cross section.
The upper boundary is chosen at the exobase level defined by condition $Kn$=1.
At this boundary we set zero
conditions for the radial derivatives of the density and temperature (so called free boundary conditions). However for the particle velocity we set the condition based on the modified Jeans escape approximation. Assuming a shifted Maxwell distribution function for the escaping particles, we derived approximate equation
matching the hydrodynamic flow below the exobase with the kinetic escape above the exobase
\begin{eqnarray}
\rho V = \sqrt{{2 k_B T}}[ n_N \sqrt{m_N} f(\lambda_N, \sqrt{m_{N}/(2 k_B T)}V)+ \nonumber  \\
n_{N2} \sqrt{m_{N2}}f(\lambda_{N2},\sqrt{m_{N2}/(2 k_B T)}V)] ,
\end{eqnarray}
where $\lambda_{N, N2}$ is the Jeans escape parameter at the exobase, i.e.
\begin{eqnarray}\label{eq:jeans}
\lambda_N = G\frac{ M_{\rm pl} m_N}{k_B T_{exo} r_{exo}}, \quad  \lambda_{N2} = G\frac{ M_{\rm pl} m_{N2}}{k_B T_{exo}r_{exo}},  \label{lambda}
\end{eqnarray}
where the function $f$ is defined as
\begin{eqnarray}
f(\lambda,u)= \frac{1}{2\sqrt{\pi}u} \int_{\sqrt{\lambda}}^{\infty}{
\left[1+(2 v u -1)\, e^{2 v u} \right ] v \,e^{-u^2-v^2}dv}.
\end{eqnarray}
In the particular case of zero velocity this expression is reduced to the well known Jeans formula
\begin{eqnarray}
f(\lambda,0) = \frac{1}{2\sqrt{\pi}} (1+ \lambda) \exp(-\lambda).
\end{eqnarray}
We calculate the upper atmosphere structures and nitrogen escape rates along EUV luminosity evolution tracks for slow, moderate and fast rotating young Sun-like stars \citep{Tu2015,Johnstone2015}.

\begin{figure}
\begin{center}
\includegraphics[width=0.95\columnwidth]{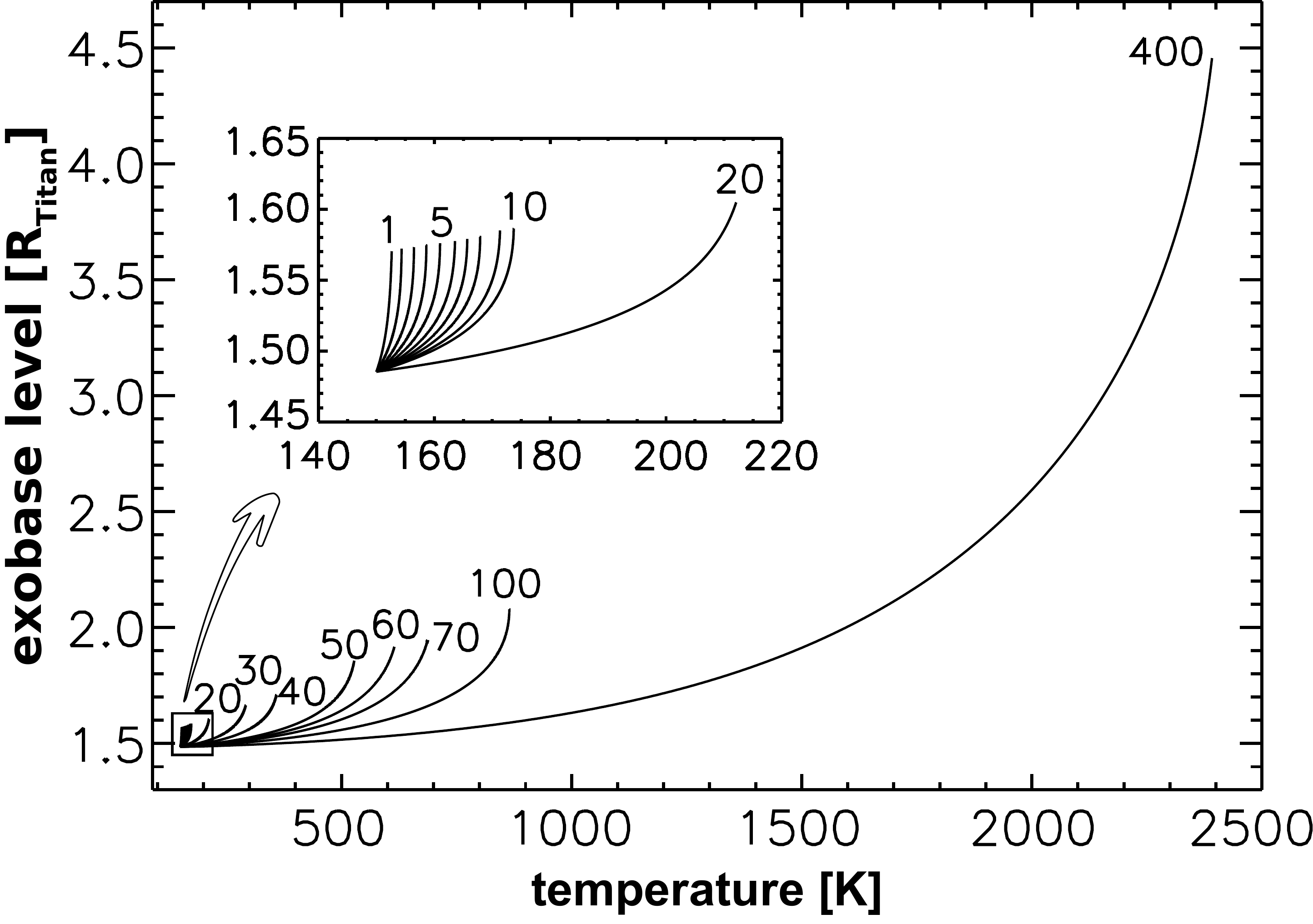}
\caption{Temperature profiles of Titan's nitrogen atmosphere according to different EUV fluxes that are normalized to the present value at Saturn's orbit. The normalized EUV flux values are given at the corresponding exobase levels
at the top of the curves.}
\label{fig:profile}
\end{center}
\end{figure}

The lower boundary of our simulation domain, \mbox{$R_{\rm 0}$}, is chosen near the homopause level at 1.485\mbox{$R_{\rm Titan}$}, where $R_{\rm Titan}$ is Titan's surface radius. We assume a nitrogen molecule number density at the lower boundary  $N_0$ = 5$\cdot$10$^8$ cm$^{-3}$.
For atmospheres that are in long-term radiative equilibrium, the temperature $T_{\rm 0}$ near the lower boundary of the simulation domain is quite close to the planetary effective temperature $T_{\rm eff}$ =150$^\circ$K.

Fig.~\ref{fig:profile} shows simulated nitrogen atmosphere temperature profiles of Titan between the present-day EUV flux exposure (EUV=1) and 400 times that of the present value (hereafter, 1\,EUV$_{\odot}$ is defined as the average present-day solar EUV flux). The curves end at the corresponding exobase levels, given in Titan radii. As one can see, for the expected EUV flux value at the time when the disk dissipated $\approx$\,4\,Myr after the origin of the solar system \citep{Wang2017,Bollard2017} the exobase level of a nitrogen-dominated upper atmosphere around Titan would have reached an exobase temperature of about 2400\,K. Already for EUV fluxes $\ge\,50\,\rm EUV_{\odot}$ the upper atmosphere is heated to more than 500\,K.

\begin{figure}
\begin{center}
\includegraphics[width=0.95\columnwidth]{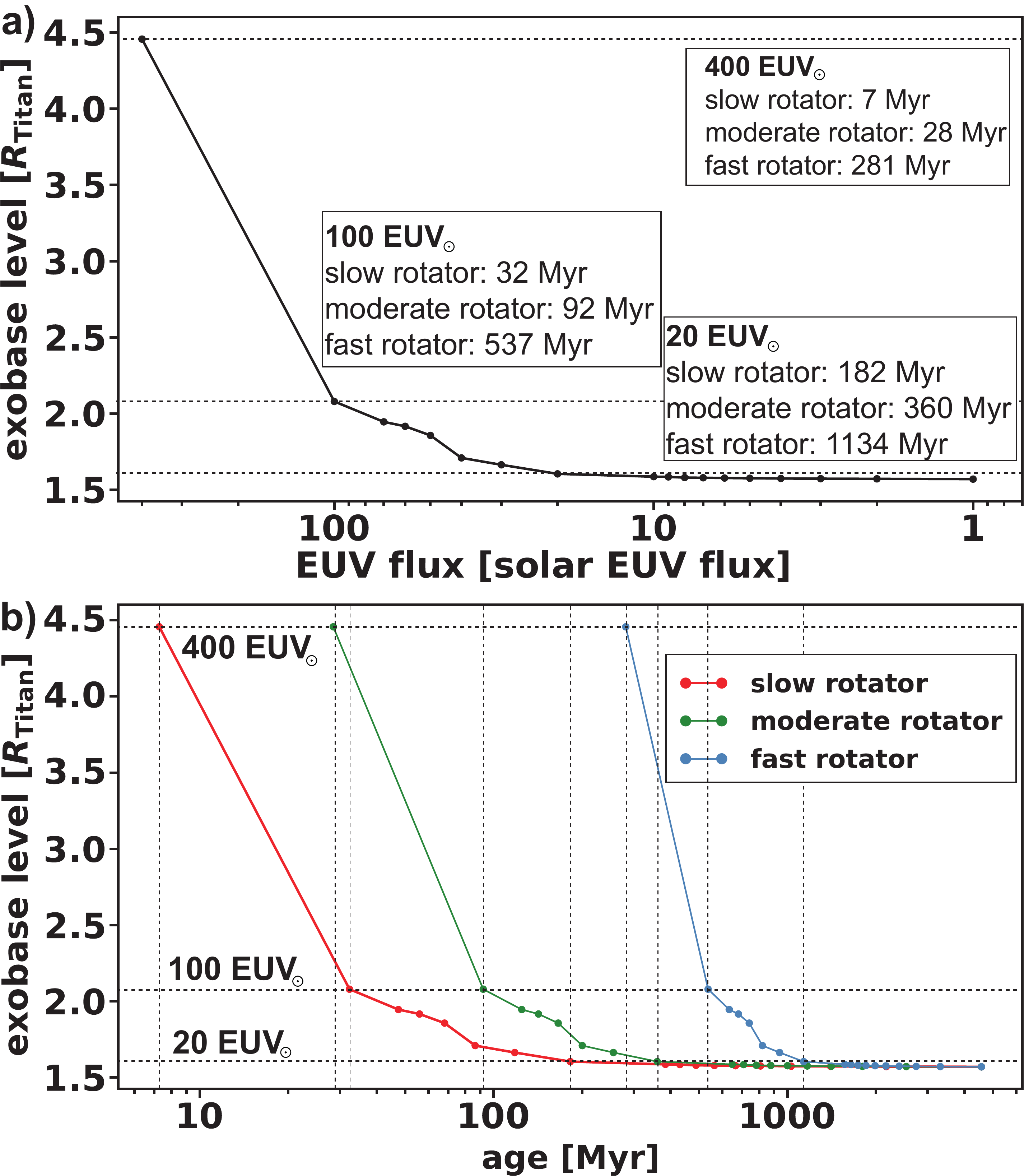}
\caption{a) Titan's exobase altitude in Titan radii as a function of EUV flux enhancement values normalized to that of the present value. b) The same but as a function of age for a slow, moderate and fast rotating young Sun.}
\label{fig:exo}
\end{center}
\end{figure}

Fig.~\ref{fig:exo}a shows the simulated exobase levels as a function of the EUV flux. One can see that the upper atmosphere does not expand to a great extent when the atmosphere is exposed to EUV fluxes that are lower than about 20\,EUV$_{\odot}$). For higher values the exobase slowly expands and reaches about 2\,R$_{\rm Titan}$ when it is exposed to 100\,EUV$_{\odot}$. For about 400\,EUV$_\odot$, the highest EUV flux expected shortly after the evaporation of the solar nebula and as long as the young Sun evolved through its activity saturation phase, the exobase level expanded to a distance of about 4.5\,R$_{\rm Titan}$. A value of 20\,EUV$_{\odot}$ corresponds to an age of the young Sun of about 4.34\, billion years ago (Ga) for a slow, and 4.2\,Ga and 3.5\,Ga for a moderate and fast rotating G-star \citep{Tu2015}.

Fig.\ref{fig:exo}b shows the exobase levels as a function of age for a slow, moderate and fast rotating Sun. The exobase levels are located at the same distances but are shifted in time due to the different rotational evolution tracks, since the EUV flux of a slow rotator decreases faster than for a moderate or even a fast rotating Sun \citep{Tu2015}. The extended upper atmospheres also result in high thermal, and even hydrodynamic escape of nitrogen. The various escape processes and loss rates are discussed in the next section.

\section{NITROGEN ESCAPE AT TITAN}

Before the arrival of Cassini at Saturn, models that were based on Voyager data obtained hydrogen escape rates from Titan's upper atmosphere in the order of
$\approx 1–3\times 10^{28}$ amu s$^{-1}$ \citep{Lebonnois2003}, and much lower escape rates for carbon and nitrogen species of $\approx 5\times 10^{26}$ amu s$^{-1}$ \citep[e.g.,][]{Gan1992,Shem2003,Cravens1997,Lammer1993,Strobel1992,Michael2005b}. In these studies the present day escape of these heavier species was dominated by atmospheric sputtering \citep{Johnson2010}. After the arrival of Cassini in the Saturn-Titan system, the descent through Titan's atmosphere of the Huygens probe with its Atmospheric Structure Instrument, and the analysis of the structure of \textbf{Titan's} thermosphere and corona inferred from its Ion and Neutral Mass Spectrometer, have also led to higher escape rate estimates for heavy species in the order of $\approx 0.3–5\times 10^{28}$ amu s$^{-1}$ \citep{Johnson2009,Johnson2010}. To understand how efficient various non-thermal atmospheric escape processes contribute to the total escape of nitrogen on present Titan is important for understanding how efficient these processes contributed to the total atmospheric escape over the satellite's history and hence for its atmospheric evolution.

In view of that, we will in the following subsections briefly discuss and summarize the loss of nitrogen through different non-thermal escape and photochemical depletion processes, before we will finally analyze thermal escape as a function of the solar EUV flux evolution.

\subsection{SUPRATHERMAL N ATOM ESCAPE}

In pre-Cassini times \citet{Lammer1991,Lammer1993} studied the atmospheric escape rates related to the production of suprathermal N($^2$D) and N($^4$S) nitrogen atoms by magnetospheric electron impact dissociation of N$_2$ molecules and dissociative recombination of N$_2^+$ ions. It was found that the main source of suprathermal nitrogen atoms originated from dissociative recombination of N$_2^+$ ions and not from electron impact dissociation of N$_2$ molecules. By using available early model based ionospheric profiles of Titan's atmosphere \citep{Keller1992,strobel1982} these authors obtained averaged nitrogen escape rates at the exobase level of $\approx 1.3 \times 10^{25}$ s$^{-1}$ and $\approx 5\times 10^{21}$ s$^{-1}$. The lower value was based on the ionosphere model of \citet{Keller1992} that included ion-neutral chemistry processes that converted the ionized major atmospheric species into H$_2$CN$^+$, HCN$^+$ and other hydrocarbon ions so that the N$_2^+$ ion density at the exobase level remained small.

New insights into Titan's ionosphere was obtained by data from several instruments onboard the Cassini Orbiter, such as the Ion and Neutral Mass Spectrometer (INMS) and the RPWS/LP (Radio and Plasma Wave – Langmuir Probe). \citet{Cravens2006} analyzed Cassini INMS measurements of ion densities of Titan and confirmed the model results of \citet{Keller1992,keller1998} that complex ion chemistry is operating and complex molecule ions such as HCNH$^+$ and C$_2$H$_5^+$ are main species. Molecular N$_2^+$ ions have a low density near the exobase level that is $< 100$ cm$^{-3}$. By using the ionospheric data analyzed after the Cassini mission one finds that suprathermal N atoms that originated from dissociative recombination escape near the exobase level with a rate of $<10^{22}$ s$^{-1}$.

Other production reactions of suprathermal atoms caused by exothermic ion and neutral chemistry in Titan's upper atmosphere were studied by \citet{DeLaHaye2007b} with a two-stream model simulation. It was found that the fraction of the absorbed energy leads to dissociations and exothermic reactions that produce atoms and molecules with energies that exceed the escape energy, so that these particles can escape to space and contribute to atmospheric escape. These authors, compared the resulting suprathermal particle fluxes, energy distributions, and hot density profiles to the N$_2$ and CH$_4$ INMS exospheric data \citep{DeLaHaye2007a}.

The contribution of exothermic ion and neutral chemistry to Titan's escape of N atoms in all forms due to exothermic chemistry are estimated by \citet{DeLaHaye2007b} to be $\approx 8.3\times 10^{24}$ s$^{-1}$. 

\subsection{PLASMA INDUCED N$_2^+$ EROSION}

Eroded ions were observed in Titan's exosphere above 1400\,km. The ion velocity profile was obtained from the difference between the observed profile and that in diffusive equilibrium, which leads to an estimated global total average ion (i.e., N$_2^+$, CH$_4^+$, H$_2^+$) escape rate between of about $1.7\times 10^{25}$ s$^{-1}$ \citep{Cui2010}. Three-dimensional multispecies hybrid simulation models that use Cassini data yield an ion loss for N$_2^+$ molecular ions in the order of about $1.3\times 10^{25}$ s$^-1$ \citep{Modolo2008,Sillanpaeae2006}.

However, the loss rates obtained by the above mentioned models are caused by the plasma interaction between the corotating magnetospheric plasma as long as Titan
is within Saturn's magnetosphere. Due to a denser and faster solar wind in the past, however, Saturn's magnetosphere was more compressed and Titan's orbit partially started to \textbf{be} outside the magnetosphere, where it was exposed to the solar wind. In such a case the plasma induced N$_2^+$ erosion inside the magnetosphere will be substituted by ion pickup escape induced through the interaction of \textbf{Titan's} atmosphere with the solar wind particles. To estimate ion pickup, we considered the same analytical approximation as \citet{Bauer1983} and \citet{Penz2005}, i.e. a momentum balance between the solar wind and the photo-ions and steady-state conditions. The mass loss rate due to ion pickup escape outside the magnetosphere can then be approximated by \citep{Penz2005}
\begin{equation}\label{pickup}
  \dot{M}_{\rm pout} \approx -2K\rho_{\rm sw}v_{\rm sw}r_{\rm exo}^2,
\end{equation}
with $K\approx0.3$ as the so-called mass loading limit \citep{Michel1971}, $\rho_{\rm sw}$ and $v_{\rm sw}$ as solar wind {\bf mass density} and velocity, and $r_{\rm exo}$ as the radius at Titan's exobase level. To calculate the values for $\rho_{\rm sw}$ and $v_{\rm sw}$ we applied the solar wind evolution model of \citet{Johnstone2015I,johnstone2015II} for a slow, moderate and fast rotating Sun. Here it is important to note that this model retrieves much lower values for $\rho_{\rm sw}$ and $v_{\rm sw}$ as were implemented in the study by \citet{Penz2005} who seem to have overestimated the ancient solar mass loss as some recent studies \citet{Johnstone2015I,johnstone2015II} indicate.

To estimate the fraction of Titan's orbit outside the Saturnian magnetosphere in the past, we applied the empirical magnetopause model of \citet{Arridge2006} which is based on magnetopause crossings of the Cassini spacecraft. Here, the polar coordinates ($r_{\rm mp}$, $\Theta$) of the magnetopause can be given as \citep{Arridge2006,Shue1997}
\begin{equation}\label{rmp}
  r_{\rm mp} = r_0\left(\frac{2}{1+\cos{\Theta}}\right)^{\kappa},
\end{equation}
where the magnetopause standoff distance $r_0$ and the exponent $\kappa$ are functions of the dynamic pressure of the solar wind $p_{\rm sw}$:
\begin{eqnarray}
  r_0 &=& a_1 p_{\rm sw}^{-a_2}  \\
  \kappa &=& a_3 + a_4 p_{\rm sw},
\end{eqnarray}
\begin{figure}
\begin{center} 
\includegraphics[width=0.95\columnwidth]{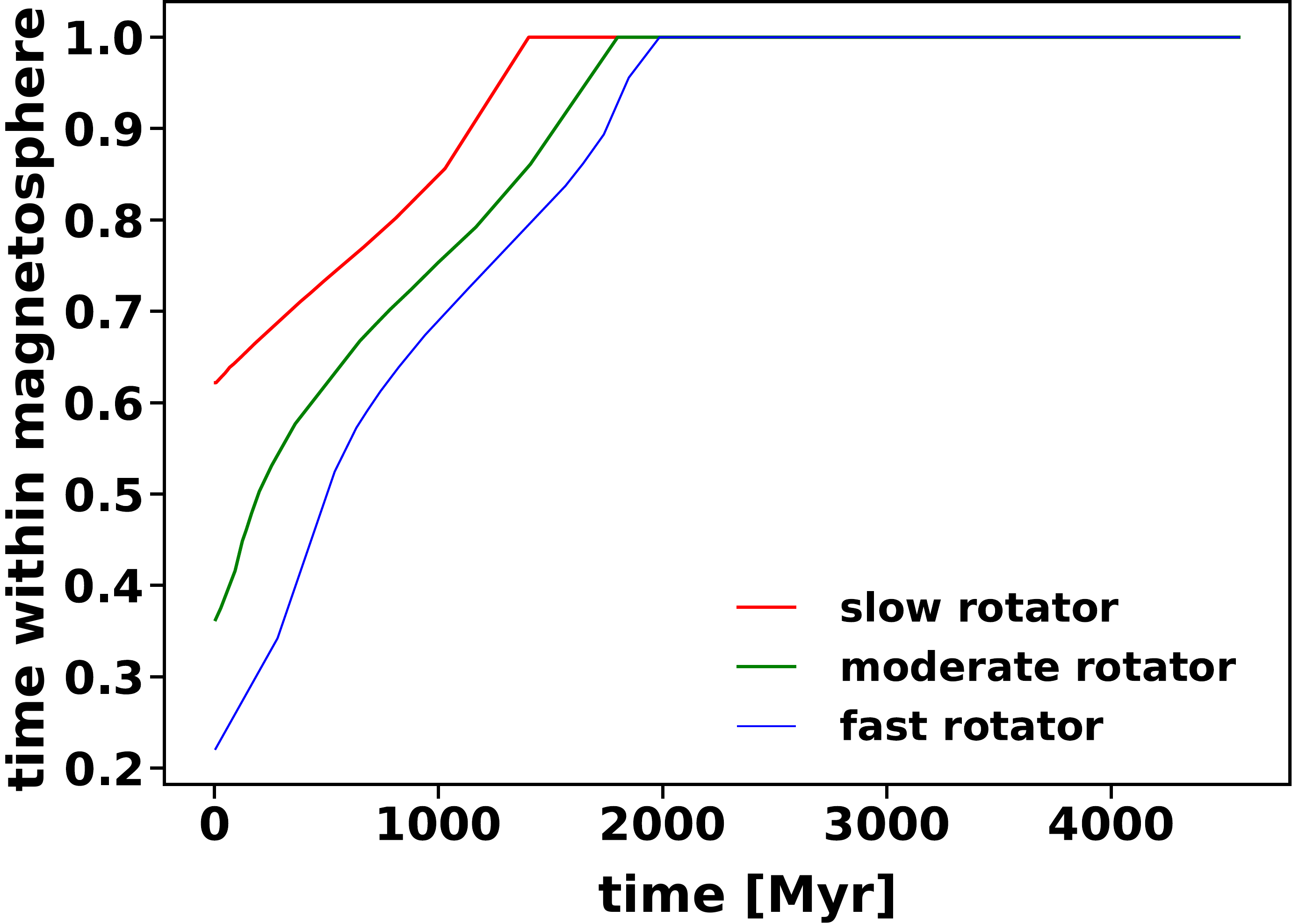}
\caption{Fraction of the time of Titan's orbit within the magnetopause of Saturn. At present-day its orbit is basically within the magnetopause for quiet solar wind conditions. Towards the past the magnetopause is more and more moving inwards due to the higher solar wind pressure and the orbit of Titan is correspondingly moving partially outwards. The effect is more pronounced for a fast than for a moderate and slow rotating Sun, respectively.}
\label{fig:mag}
\end{center}
\end{figure}
with the fitted coefficients $a_1=9.7\pm 1.0$, $a_2=0.24\pm 0.02$, $a_3=0.77\pm 0.03$, and $a_4=-1.5\pm 0.3$ based on magnetopause crossings of the Cassini spacecraft \citep{Arridge2006}. Fig.~\ref{fig:mag} illustrates the fraction of time $t_{\rm in}$ of Titan's orbit that is within the Saturnian magnetosphere for a slow, moderate and fast rotating Sun. While at present-day the orbit of Titan is completely within the magnetopause for quiet solar wind conditions it starts significantly being outwards earlier in the past. For such cases we therefore assume plasma induced N$_2^+$ erosion inside and ion pickup outside the magnetosphere according to the respective fraction of Titan's orbit within and outside the magnetopause, i.e. the total plasma induced ion loss can then be written as
\begin{equation}\label{ionloss}
  \dot{M}_{\rm ptot} = \dot{M}_{\rm pin}\times t_{\rm in} + \dot{M}_{\rm pout}\times (1-t_{\rm in}),
\end{equation}
with $\dot{M}_{\rm pin}=1.5\times10^{25}\mathrm{s}^{-1}$ as plasma induced $N_2^+$ erosion within the magnetosphere which we assumed to be constant over time.

It has further, however, to be noted that the evolution of Saturn's magnetosphere might not be that straight forward. At present time, Saturn is expected to have a helium-rich layer in its interior on top of the \textbf{convective layer} that generates the magnetic field. This layer is produced by helium rain due to separation of helium from hydrogen early-on in the history of the giant planet at a time when its hydrogen-dominated atmosphere cooled to a certain temperature, thereby making helium immiscible with hydrogen for the respective pressure range \citep[e.g.,][]{Stevenson1977,Stevenson1980,Fortney2003,Puestow2016}. Since such a layer dampens the surface magnetic field of Saturn, its magnetosphere is currently smaller than scaling laws expect for such a planet \citep{Stevenson1980,Christensen2006}. \citet{Puestow2016} found that the helium-rich layer should have formed not earlier than about one billion years after the formation of Saturn for an atmospheric temperature of 183\,K. It can, therefore, be speculated that this effect counteracted the higher solar wind ram pressure as long as the helium-rich layer did not build up. This might have increased the Saturnian magnetosphere, thereby keeping Titan potentially inside.

\subsection{ATMOSPHERIC SPUTTERING OF N ATOMS}

The third non-thermal atmospheric escape process is sputtering caused by i) the magnetospheric plasma flow when the satellite is inside or by the solar wind plasma when it is outside Saturn's magnetosphere, and ii) by back-scattered pickup ions from Titan's exosphere. \citet{Lammer1993} applied an analytical simple sputter model to Titan's upper atmosphere and estimated N atom escape rates in the order of $3-7\times 10^{25}$ s$^{-1}$, depending on whether Titan was exposed to the solar wind or the magnetospheric corotating plasma flow. \citet{Shem2003} applied a more sophisticated 1-D Monte Carlo model and hybrid (particle ions, fluid electrons) plasma flow model results from simulations of Saturn's magnetospheric interaction with Titan's exosphere \citep{Brecht2000}, and obtained average sputter escape rates of $\approx 3.6\times 10^{25}$ s$^{-1}$. \citet{Michael2005} used a 3-D Monte Carlo model for studying the sputtering efficiency of N atoms and N$_2$ molecules from Titan's upper atmosphere due to the interaction of Saturn's magnetospheric ions and molecular pick-up ions. These authors obtained total nitrogen escape rates from Titan's upper atmosphere when exposed to N$^+$ and N$_2^+$ ions from the magnetospheric plasma flow of $\approx 1.2\times 10^{25}$ s$^{-1}$ (N atoms: $\approx 8.8\times 10^{24}$ s$^{-1}$; N$_2$ molecules: $\approx 1.4\times 10^{24}$ s$^{-1}$) and $\approx 2.5\times 10^{25}$ s$^{-1}$ (N atoms: $\approx 1.6\times 10^{25}$ s$^{-1}$; N$_2$ molecules: $\approx 4.6\times 10^{24}$ s$^{-1}$), respectively. Thus, the total calculated nitrogen escape rate caused by magnetospheric plasma induced sputtering is of the order of $\approx 3.7\times 10^{25}$ s$^{-1}$.

For sputtering the same situation applies as for plasma induced N$_2^+$ erosion, i.e. if Titan's position is outside the Saturnian magnetosphere, solar wind induced sputtering dominates over magnetospheric plasma induced sputtering. Outside the magnetopause we therefore also apply an approximation that is based on \citet{Johnson1990} and \citet{Penz2005}, i.e.
\begin{equation}\label{sputtout}
  S_{\rm out} = \rho_{\rm sw}v_{\rm sw}2\pi r_{\rm exo}^2\left(\frac{0.5\sigma_{d,a}}{\sigma_d\cos{\Phi}}+\frac{3\alpha S_{n}}{\pi^2 (cos{\Phi}^{1.6} E_{\inf}\sigma_d}\right),
\end{equation}
with $\Phi$ as pitch angle of the incident particles, $\sigma_{d,a}=3\times10^{-15}$\,cm$^2$ as collision cross-section for a particle receiving an energy transfer, $\sigma_d=1.667\times 10^{-15}$\,cm$^1$ as cross-section for the escape of the struck particle, $S_n=10^{-14}$\,eV\,cm$^2$ as stopping cross-section, $E_{\inf}=0.036$\,ev/amu as escape energy for particles at the exobase \citep{Lammer1993}, and $\alpha=0.8$ as a factor that depends on the mass ratio $m_a/m_b$ of the target particle mass $m_b$ to the incident particle mass $m_a$.

\subsection{PHOTOCHEMICAL NITROGEN DEPLETION PROCESSES}

Besides atmospheric escape processes, N$_2$ can also be depleted in Titan's atmosphere through photodissociation of N$_2$ into 2\,N and subsequent incorporation into nitriles such as HCN \citep{Liang2007,Krasnopolsky2016}. Once a nitrogen atom gets incorporated into a nitrile it will be inevitably lost from the atmosphere due to the triple bond between the C and N atom through either condensation or polymerization of the nitriles. The present-day loss rate of N$_2$ due to these processes was recently estimated to be $ 9.7 \times 10^{22} $s$^{-1}$ \citep{Krasnopolsky2016}. To incorporate nitrogen into nitriles the presence of CH$_4$ in the atmosphere is needed to form nitriles from both dissociation products, i.e. without CH$_4$ this depletion process would halt. It is therefore important to consider the evolution and availability of CH$_4$ in the atmosphere of Titan during its past. Interior models \citep{Tobie2006,Tobie2012,CastilloRogez2010} suggest that outgassing of CH$_4$ happened relatively late in the history of Titan, most probably mainly within the last 1\,Gyr, even though episodic outgassing might have also occurred earlier \citep{Tobie2006}. In addition, the $^{12}$C/$^{13}$C ratio of 88.5\,$\pm$\,1.4 in CH$_4$ \citep{Mandt2012} is also suggesting methane outgassing during the last 500\,Myr, but not longer than about 1\,Gyr. However, there could have also been extended periods of episodic methane outgassing prior to 1\,Gyr ago as interior models suggest \citep{Tobie2006}. Even though these potential periods of outgassing could as well have been drivers for photochemical removal, the fractionation potential of this process might have been significantly lower at earlier times since it is anticorrelated with the intensity of the EUV flux as research on its variation over the solar cycle indicates \citep{Mandt2017}. Since it is whether \textbf{not} known how frequently and in what amounts CH$_4$ was outgassed early on, nor how significantly higher EUV fluxes reduce the efficiency of this potential fractionation process, we assume photochemical sequestration of N$_2$ into nitriles to be a relevant loss and fractionation process not longer than during the last 1\,Gyr, for which we assume it to be constant at $9.7 \times 10^{22} $s$^{-1}$.

\subsection{THERMAL NITROGEN ESCAPE}

Thermal atmospheric escape is confined by gravity and temperature at the exobase level and can be characterized by the Jeans parameter $\lambda$, which is the ratio of the gravitational energy to the thermal energy (see Equation~\ref{eq:jeans}).
In the absence of an atmospheric bulk flow, thermal escape can be described by the classical Jeans escape formula \citep[e.g.][]{Chamberlain1963,Bauer2004}
\begin{equation}
F_{\rm Jeans}=\frac{n_{\rm exo} \widetilde{v}}{4} (1+\lambda)e^{(-\lambda)},
\end{equation}
with $\widetilde{v}$ as the mean thermal velocity and $n_{\rm exo}$ as the exobase density.
If $\lambda$ is less than the critical value of about $\lambda \approx 2.5$, a stationary hydrodynamic transition from subsonic to supersonic flow \citep{Volkov2011,Volkov2013,Erkaev2015} is not possible. In such a case the atmosphere is no longer bound and fast non-stationary atmospheric expansion can occur that results in extreme thermal atmospheric escape.
For larger values of the Jeans parameter exceeding the critical value, a stationary atmospheric bulk flow can be driven by the upper atmosphere heating caused by the solar EUV radiation. The hydrodynamic escape regime can be realized when the bulk velocity reaches the supersonic value at the exobase.
Otherwise, if the bulk velocity is still subsonic at the exobase, the slow hydrodynamic flow below the exobase should be matched with the kinetic escape above the exobase.

\citet{Strobel2008,Strobel2009} suggested a so-called ``slow hydrodynamic escape model'' which yielded grossly overestimated N$_2$ escape rates from Titan's upper atmosphere of $\approx 4 - 5 \times 10^{28}$\,amu\,s$^{-1}$. This model requires an extended quasi-collisional region above the exobase level where efficient energy transfer could occur. As described in Sect.~\ref{Sec2}, we applied 1D radial continuity, momentum and energy equations that are scaled by the escape flux, $\lambda$ and temperature. The flux of molecules at the exobase level with velocities above the escape velocity is much too small to account for the extracted escape flux.  \citet{Strobel2008} and \citet{Strobel2009} continued the modelling above the exobase level where the energy is assumed to be supplied by upward thermal conduction.

However, the fluid expression used for conduction fails as the Knudsen number increases above about 0.1, and fails above the exobase \citep{Tucker2009,Schaufelberger2012}. The results of kinetic models show that the collision probability of nitrogen and methane above Titan's exobase level decreases very fast with height \citep{Schaufelberger2012}, thereby not supporting the existence of an extended quasi-collisional region above the exobase level, necessary for the physically inaccurate model approach of \citet{Strobel2008,Strobel2009}. From the brief discussion above, one can conclude that thermal escape will not contribute to the observed escape rates at present-day Titan to a large extent. This is also obtained by our upper atmosphere model results, from which we retrieve thermal nitrogen escape rates of about $ 7.5 \times 10^{17}$ s$^{-1}$ for the present-day. However, for higher EUV fluxes, as we will see in the following section, thermal escape can become significant.

\citet{Johnson2013} developed a criterion for estimating when for thermal escape the atmospheric outflow goes transonic in the continuum region. Their criterion can be used to investigate on whether subsonic or transonic solutions should be implemented into an upper atmosphere model for certain environmental conditions.  If the integrated heating and cooling rate $Q_{\mathrm{net}}$ exceeds the so-called sonic boundary condition $Q_c$, transonic solutions should be applied and vice verca. Here, $Q_{\mathrm{net}}$ is defined as \citep{Johnson2013}
\begin{equation}\label{Eq:Qnet}
  Q_{\rm net} \approx 4\pi r^2 n M a \sqrt{\frac{\gamma}{\lambda}\frac{U(r)}{\mu}}U(r_0),
\end{equation}
where $r$ is the radius at which $Q_{\rm net}$ is estimated, $Ma$ is the Mach number, $\gamma$ is the heat capacity, $U(r)$ and $U(r_0)$ are the gravitational energies at $r$ and the lower boundary of the simulation region $r_0$, respectively. \citet{Johnson2013}, then define their criterion as
\begin{equation}\label{Eq:criterion}
  Q_{\rm net} > Q_C \approx 4 \pi r_{\rm exo} \frac{\gamma}{c_c \sigma_c Kn_m}\sqrt{\frac{2 U (r_{\rm exo})}{m}}U(r_0),
\end{equation}
where $c_c$ is $\sqrt{2}$, and $Kn_m$ is $\approx 1$ if the heat is mainly absorbed over a broad range of $r$ below $r_{\rm exo}$. If we apply this criterion to our model for the thermal escape at 400\,EUV$_{\odot}$, we retrieve $Q_{\rm net}/Q_c = 1.3$, which is in agreement with our transonic solution for this EUV flux. In case of 100\,EUV$_{\odot}$ we have a subsonic outflow, for which we use a modified Jeans boundary condition at the exobase boundary. In this case, \textbf{we retrieve $Q_{\rm net} = 7.7 \times 10^{9}$\,W and $Q_c = 1.p5 \times 10^{10}$\,W,} which equals $Q_{\rm net}/Q_c = 0.5$. This is again in agreement with the criterion of \citet{Johnson2013}.

\subsection{NITROGEN ESCAPE THROUGH TIME}
\begin{figure}
\begin{center} 
\includegraphics[width=0.95\columnwidth]{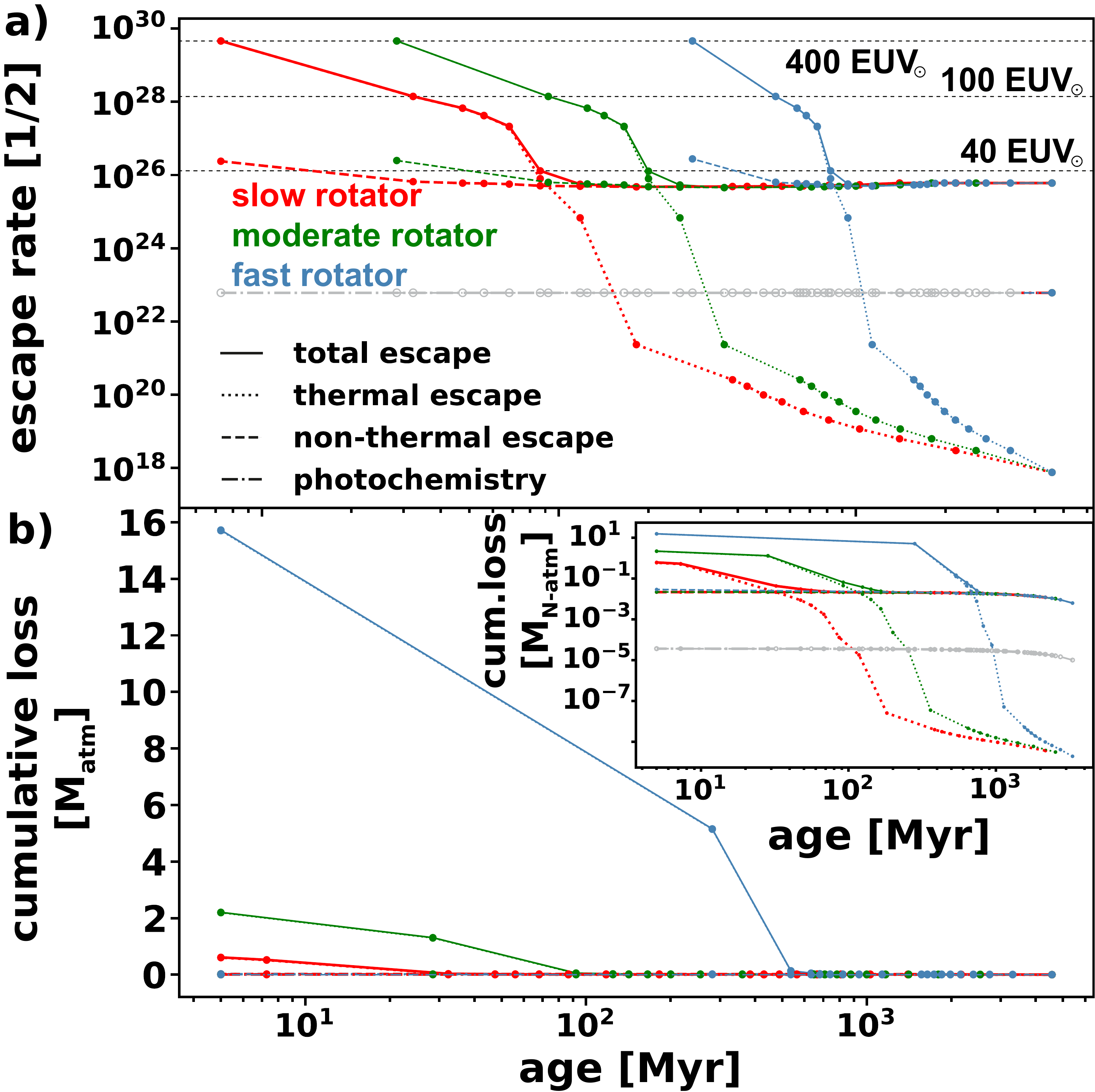}
\caption{a) Thermal (dotted) and non-thermal escape (dashed), as well as photochemical loss of nitrogen from the atmosphere (dash-dotted) for a slow (red), moderate (green) and fast rotator (blue) over time. The solid line shows the sum over all loss processes. For photochemical loss only the last 1000\,Myr were considered (colored part; non-considered part illustrated in gray). b) Cumulative loss over time in Titan-atmospheres for the same rotational evolutions of the Sun. The insert shows the cumulative loss over time with logarithmic y-axes to illustrate the separate non-thermal and thermal losses in more detail. Here, the figure starts with EUV$_{\odot}\,=2\,.$}
\label{fig:escape}
\end{center}
\end{figure}

\begin{center}
\begin{table*}
\begin{threeparttable}
\caption{Loss of nitrogen atoms [$\mathrm{s^{-1}}$] through different escape processes for 1, 20, 100 and 400 EUV for a slow, moderate, and fast rotating young Sun. The last column shows the total, thermal, non-thermal and photochemical (for 1\,Gyr) loss of nitrogen (present-day Titan-atmospheres [$\rm M_{atm}$]) if the atmosphere would have existed from the beginning of the solar system. \textbf{The abbreviation ``i.m.'' means inside the magnetosphere, ``o.m.'' outside the magnetosphere.}}
\begin{tabular}[t]{l l c c c c c} \hline
 & & 1 EUV & 20 EUV & 100 EUV & 400 EUV & total loss \\ \hline
 \textbf{slow} & \textbf{total} & $6.03\times 10^{25}$ & $4.76 \times 10^{25}$ & $1.41 \times 10^{28}$ & $4.60 \times 10^{29}$ & 0.62 \\ \hline
 & thermal & $7.50 \times 10^{17}$ & $2.32 \times 10^{21}$ & $1.40 \times 10^{28}$ & $4.60 \times 10^{29}$ & 0.60 \\ \hline
 & non-thermal (total)$^a$ & $6.03 \times 10^{25}$ & $4.64 \times 10^{25}$ & $4.83 \times 10^{25}$ & $7.57 \times 10^{25}$ & 0.02 \\
 & \textit{sputtering i.m.} & $3.70 \times 10^{25}$ & $3.70 \times 10^{25}$ & $3.70 \times 10^{25}$ & $3.70 \times 10^{25}$  & \\
 & \textit{sputtering o.m.} & $2.61 \times 10^{24b}$ & $9.80 \times 10^{24}$ & $1.84 \times 10^{25}$ & $8.63 \times 10^{25}$ & \\
 & \textit{ion escape i.m.} & $1.50 \times 10^{25}$ & $1.50 \times 10^{25}$ & $1.50 \times 10^{25}$ & $1.50 \times 10^{25}$ & \\
 & \textit{ion escape o.m.} & $1.96 \times 10^{23b}$ & $7.35 \times 10^{23}$ & $1.38 \times 10^{24}$ & $6.48 \times 10^{24}$ & \\
 & \textit{suprathermal} & $8.3 \times 10^{24}$ & $8.3 \times 10^{24}$ & $8.3 \times 10^{24}$ & $8.3 \times 10^{24}$ & \\ \hline
 & photochemistry & $9.7 \times 10^{22}$ & -- & -- & -- & $8.05\times10^{-5}$ \\ \hline\hline
  \textbf{moderate} & \textbf{total} & $6.03 \times 10^{25}$ & $4.53 \times 10^{25}$ & $1.41 \times 10^{28}$ & $4.60 \times 10^{29}$ & 2.22 \\ \hline
 & thermal & $7.50 \times 10^{17}$ & $2.32 \times 10^{21}$ & $1.40 \times 10^{28}$ & $4.60 \times 10^{29}$ & 2.20 \\ \hline
 & non-thermal (total)$^a$ & $6.03 \times 10^{25}$ & $4.43 \times 10^{25}$ & $5.17 \times 10^{25}$ & $1.53 \times 10^{26}$ & 0.02 \\
 & \textit{sputtering i.m.} & $3.70 \times 10^{25}$ & $3.70 \times 10^{25}$ & $3.70 \times 10^{25}$ & $3.70 \times 10^{25}$  & \\
 & \textit{sputtering o.m.} & $2.61 \times 10^{24b}$ & $1.32 \times 10^{25}$ & $3.46 \times 10^{25}$ & $1.86 \times 10^{26}$ & \\
 & \textit{ion escape i.m.} & $1.50 \times 10^{25}$ & $1.50 \times 10^{25}$ & $1.50 \times 10^{25}$ & $1.50 \times 10^{25}$ & \\
 & \textit{ion escape o.m.} & $1.96 \times 10^{23b}$ & $9.90 \times 10^{23}$ & $2.60 \times 10^{24}$ & $1.39 \times 10^{25}$ & \\
 & \textit{suprathermal} & $8.3 \times 10^{24}$ & $8.3 \times 10^{24}$ & $8.3 \times 10^{24}$ & $8.3 \times 10^{24}$ & \\ \hline
 & photochemistry & $9.7 \times 10^{22}$ & -- & -- & -- & $8.05\times10^{-5}$ \\ \hline\hline
 \textbf{fast} & \textbf{total} & $6.03 \times 10^{25}$ & $4.53 \times 10^{25}$ & $1.41 \times 10^{28}$ & $4.60 \times 10^{29}$ & 15.73 \\ \hline
 & thermal & $7.50 \times 10^{17}$ & $2.32 \times 10^{21}$ & $1.40 \times 10^{28}$ & $4.60 \times 10^{29}$ & 15.70 \\ \hline
 & non-thermal (total)$^a$ & $6.03 \times 10^{25}$ & $4.85 \times 10^{25}$ & $4.90 \times 10^{25}$ & $1.90 \times 10^{26}$ & 0.03 \\
 & \textit{sputtering i.m.} & $3.70 \times 10^{25}$ & $3.70 \times 10^{25}$ & $3.70 \times 10^{25}$ & $3.70 \times 10^{25}$  & \\
 & \textit{sputtering o.m.} & $2.61 \times 10^{24b}$ & $8.43 \times 10^{24}$ & $2.63 \times 10^{25}$ & $2.32 \times 10^{26}$ & \\
 & \textit{ion escape i.m.} & $1.50 \times 10^{25}$ & $1.50 \times 10^{25}$ & $1.50 \times 10^{25}$ & $1.50 \times 10^{25}$ & \\
 & \textit{ion escape o.m.} & $1.96 \times 10^{23b}$ & $6.33 \times 10^{23}$ & $1.97 \times 10^{24}$ & $1.74 \times 10^{25}$ & \\
 & \textit{suprathermal} & $8.3 \times 10^{24}$ & $8.3 \times 10^{24}$ & $8.3 \times 10^{24}$ & $8.3 \times 10^{24}$ & \\ \hline
 & photochemistry & $9.7 \times 10^{22}$ & -- & -- & -- & $8.05\times10^{-5}$ \\ \hline
\label{tab:escape}
\end{tabular}
\begin{tablenotes}
\item $^a$\footnotesize{Average value over one orbit around Saturn;}
\item $^b$\footnotesize{Hypothetical value if Titan would be outside the magnetosphere at 1 EUV.}\\
\end{tablenotes}
\end{threeparttable}
\end{table*}
\end{center}
We simulated thermal and extrapolated non-thermal escape and photochemical loss of nitrogen through time for a slow, moderate and fast rotating Sun. Fig.~\ref{fig:escape}a shows the evolving nitrogen escape rates of these various escape processes during time as a function of solar age-related EUV flux for a slow (red curves), moderate (green curves) and fast rotator (blue curves), respectively. Here, `non-thermal escape' depicts the sum over all non-thermal escape processes, that is, ion escape and sputtering inside and outside the Saturnian magnetosphere, and suprathermal escape. Our results show that for solar EUV fluxes $>\,30\mathrm{EUV_{\odot}}$ thermal escape rates become larger than all before mentioned non-thermal escape rates together. From our simulation we obtain EUV flux dependent thermal nitrogen escape rates between $ 7.5 \times 10^{17}$ for 1\,EUV$_{\odot}$ and 4.6\,$\times 10^{29}$ s$^{-1}$ for 400\,EUV$_{\odot}$ (see also Table~\ref{tab:escape}). This illustrates that the present-day thermal nitrogen escape rates from Titan are negligible compared to the present day losses caused by non-thermal escape processes such as the ambient plasma flow, pick-up ions, energetic re-impacting neutrals and related atmospheric sputtering, as well as loss through suprathermal atoms \citep[see e.g.][for estimates on non-thermal escape processes at present-day Titan]{Lammer1991,Lammer1993,Shem2003,Michael2005,Michael2005b,Penz2005,DeLaHaye2007a,DeLaHaye2007b,Mandt2009,Mandt2014}. In the distant past, however, the picture changes completely and it also gets crucially important whether the Sun was a slow, moderate or fast rotator.

Our result of thermal escape for 400\,EUV$_{\odot}$ is further in agreement with simulated escape rates for early Titan by \citet{Johnson2016}, who retrieved escape rates of $1.7 \times 10^{30}$\,s$^{-1}$ for $Q_{\rm net} = 13.2 \times 10^{10}$\,W and $T_0 \approx 150$\,K, whereas our value is $4.6 \times 10^{29}$\,s$^{29}$\,s$^{-1}$ for $Q_{\rm net} = 5.3 \times 10^{10}$\,W. If we rescale our loss rate to the higher $Q_{\rm net}$, we get $1.2 \times 10^{30}$\,s$^{-1}$, which is very close to the result of \citet{Johnson2016}.

For average present-day solar wind conditions, the solar wind induced escape rate of $\rm N_2^+$ ions from Titan would be about $2 \times10^{23} {\rm s^{-1}}$ as estimated via Equation~\ref{ionloss}, which is about two orders of magnitude lower compared to the current loss inside Saturn's magnetosphere. For a slow rotating Sun, solar wind induced ion escape will stay below the ion escape rate within the magnetosphere (see Table~\ref{tab:escape}) for the whole history of the solar system. For a moderate and fast rotator, however, it will start to get comparable in strength for very early times, with $\approx1.4\times10^{25}\rm s^{-1}$ and $\approx1.7\times10^{25}\rm s^{-1}$, respectively, at 400\,$\rm EUV_{\odot}$. This is due to a stronger increase towards the past in solar wind velocity and density for a moderate and fast rotator than for a slow rotator which, in addition, also increases Titan's fraction of time outside the Saturnian magnetosphere.

Contrary to ion escape, sputtering gets stronger in the past if Titan resides outside the magnetospheric environment of Saturn for all three rotational tracks. For 400\,$\rm EUV_{\odot}$ sputtering outside the magnetosphere ranges from $8.6 \times 10^{25}\rm s^{-1}$ for a slow to $2.32 \times 10^{26}\rm s^{-1}$ for a fast rotator, while sputtering inside the magnetosphere is assumed to be constant at $3.7 \times 10^{25}\rm s^{-1}$ for all different rotators.

It has to be noted, however, even though we assumed the escape processes within the magnetosphere, i.e. sputtering, plasma induced erosion of N$_2^+$ and suprathermal escape, to be constant over time, this might not have been the case. Suprathermal escape for instance might increase back in time due to the increase of EUV flux \citep[see e.g. the recent study of ][]{Amerstorfer2017} but might on the other hand also decrease at higher EUV fluxes due to an enhanced probability of collision between energetic and thermal atoms in an expanded upper atmosphere \citep[see e.g. another recent study on Mars by ][]{Zhao2017}. Also, the complex ion chemistry that produces suprathermal nitrogen atoms should have been different in the past, since CH$_4$ likely was not available in similar abundances than at present-day over the whole history of Titan. Clearly, a study on suprathermal escape on early Titan would be needed to answer this uncertainty. In addition, plasma induced erosion and sputtering within the magnetosphere should have also changed since the Saturnian plasma environment should have likely been different within an early compressed magnetosphere and the exobase levels of Titan's atmosphere were different than today. Furthermore, Saturn could have also had a plasma torus, comparable to the present-day Jovian plasma torus, due to the enhanced loss rates from Titan's nitrogen atmosphere. Assuming these processes constant does not change our results significantly since thermal escape becomes dominant by orders of magnitude for fluxes $>30\,\rm EUV_{\odot}$, making non-thermal escape negligible. It is unlikely that these non-thermal escape processes within the magnetosphere could deviate from the assumed values in such a way that they would be as important than thermal escape in the past. Similarly, a Titan that would be more frequently than assumed in our simulations inside a bigger Saturnian magnetosphere during the first $\approx$\,1 billion years due to the helium-rich layer above the deep dynamo not yet existing, would not change our overall result substantially.

Besides non-thermal and thermal escape, we also considered photochemical sequestration of N$_2$ into nitriles as described above for the last recent 1000\,Myr with a constant rate of $9.7\times 10^{22}\rm\,s^{-1}$ \citep{Krasnopolsky2016}. For the total escape, however, this process is negligible, also if one assumes photochemical sequestration being active over the whole duration of 4.5\,Gyr. This is also illustrated clearly, if one considers the cumulative loss over time, which is $8.1 \times 10^{-5}$ Titan-atmospheres ($\rm M_{atm}$) for 1000\,Myr and $3.7 \times 10^{-4}\rm M_{atm}$ for a duration of $\approx\,$4.5\,Gyr.

For the total integrated loss over time, the most important factor -- especially during the first 100 to 1000\,Myr -- is the EUV flux evolution of the Sun. As can be seen in Figures~\ref{fig:exo} and \ref{fig:escape}, the EUV flux of a slowly rotating young Sun declines significantly at very early stages, leaving the EUV saturation level after about 5\,Myr \citep{Tu2015} and falling below $400\,\rm EUV_{\odot}$ at 7\,Myr and $100\,\rm EUV_{\odot}$ at 32\,Myr. After about 180\,Myr the EUV flux already declined to $20\,\rm EUV_{\odot}$ and the expansion of Titan's atmosphere decreased from about $4.5\,\rm R_{\rm Titan}$ at $400\,\rm EUV_{\odot}$ to just above $1.6\,\rm R_{\rm Titan}$. Consequently, the thermal escape decreased from $4.6\times10^{29}\,\rm s^{-1}$ to $2.32\times10^{21}\,\rm s^{-1}$ by 8 orders of magnitude. For a moderate and, most notably, for a fast rotating young Sun, however, thermal escape stays dominant significantly longer with $400\,\rm EUV_{\odot}$ being reached at about 30\,Myr and 280\,Myr, and $20\,\rm EUV_{\odot}$ at 360\,Myr and 1134\,Myr, respectively. For a slow rotator, thermal escape will therefore be the dominant process at Titan for only about 100\,Myr, whereas for a fast rotator it will stay dominant for more than 1\,Gyr, making a significant difference in terms of cumulative total loss over time.

Fig.~\ref{fig:escape}b shows the potential cumulative loss of nitrogen (in $\rm M_{atm}$) for a slow, moderate and fast rotator over time from present-day back until 5\,Myr after formation of the solar system, implicitly assuming that Titan's nitrogen atmosphere existed from the beginning. The solid lines show the sum over all escape processes which is almost identical with the loss through thermal escape (dotted lines, not distinguishable in this figure from the total cumulative loss). This plot clearly illustrates the importance of i) thermal escape in the past, and of ii) the EUV flux evolution on the thermal escape and, thus, on the cumulative loss over time. For a slow rotating Sun the total cumulative loss over time would be 0.62\,$\rm M_{atm}$ (of which 0.6\,$\rm M_{atm}$ are lost through thermal and 0.02\,$\rm M_{atm}$ (see also Table~\ref{tab:escape}) through non-thermal escape mechanisms, if Titan's nitrogen atmosphere would have originated at the beginning of the solar system together with the satellite itself. If the Sun originated as a moderate or fast rotating young G-type star, on the other hand, the total loss over the whole solar system lifetime would have been significantly higher, i.e. 2.20\,$\rm M_{atm}$ for a slow and 15.73\,$\rm M_{atm}$ for a fast rotating Sun.

These integrated losses, however, illustrate scenarios in which Titan's atmosphere already originated during or right after the formation of the satellite and in which the nitrogen reservoir at Titan was sufficiently big to survive the early phase of strong thermal escape. In the scenario of a fast rotating Sun for instance, Titan's atmosphere must have had an initial reservoir of 16.73\,$\rm M_{atm}$, otherwise its atmosphere would not have survived the first billion years, or in other words, Titan would not have been able to sustain a thick nitrogen-dominated atmosphere until the present-day. But our integrated losses might only be lower values if its atmosphere would have formed early-on, since we did not consider atmospheric escape through impact erosion. \citet{Marounina2015} simulated this process on Titan during the Late Heavy Bombardment (LHB) and found that it could have eroded up to 5 times the present-day atmosphere, if it was already present at that time. This would increase our estimates to initial reservoirs of 6.62, 8.20, and 21.73 times the present-day atmospheric reservoir for a small, moderate, and fast rotator, respectively. Here, it has also to be pointed out, however, that the LHB might not have taken place as a peak in impactor flux, but only as an exponentially declining tail, as recent studies indicate \citep[e.g.,][]{Boehnke2016,Morbidelli2018,Mojzsis2019}. If so, this would not exclude earlier impact erosion of Titan's atmosphere.

If Titan had collected such high nitrogen amounts that are ``theoretically'' possible to lose during the satellites lifetime for a fast rotating young G-type star, then its atmosphere should have originated directly from the solar nebula or from Saturn's subnebula. If Titan's atmosphere would have had such an exogenous origin then the present remaining fraction would contain slightly more Ne than N, while Ar/N would equal 1/30 according to the present-day solar abundances \citep{Grevesse2005,Owen2009}. Today at Titan one finds a Ne/N ratio that is $<10$ and an Ar/N ratio that is about $1.5\times 10^{-7}$ \citep{Niemann2005}. Because of these discoveries during the Cassini-Huygens mission it seems to be clear that Titan's atmosphere is secondary, and originated endogenically due to degassing from the icy building blocks that formed the satellite.

Another interesting theory, that might be connected to strong atmospheric escape, is that Titan captured a small amount of Ne from the solar nebula \citep{Glein2017}. If the $^{22}$Ne/$^{36}$Ar ratio of (0.05\,--\,2.5) times solar is presently close to the lower range, \citet{Glein2017} concluded, then up to 90\% of Titan's Ne could have been lost through atmospheric escape. If it is higher, however, this might be an implication for Ne outgassing being significantly more efficient than Ar, and for a chondritic origin of Ne \citep{Glein2015,Glein2017,Tobie2012}. Both, a nebular and a chondritic origin of Ne, would further also be in agreement with a geochemically active Titan \citep{Tobie2012,Glein2015}, and a differentiated core \citep{Baland2014,ORourke2014}.

To better understand the origin and evolution of Titan's thick nitrogen-dominated atmosphere it is therefore crucial to investigate i) the time of its origin and potential outgassing, ii) the initial reservoir out of which its nitrogen -- but also its noble gases -- originated, and iii) whether the Sun was a slow, moderate or fast rotator, or something in-between. We will address these points within the following sections, starting with potential nitrogen reservoirs and the therewith connected initial fractionation of $\rm ^{14}N/^{15}N$ in Titan's atmosphere.
\newline

\section{IMPLICATIONS AND DISCUSSION}

\subsection{FRACTIONATION OF TITAN'S NITROGEN}\label{sec:frac}

The isotopic ratio of $^{14}$N/$^{15}$N in Titan's atmosphere can tell us something about the building blocks of its atmosphere since the potential reservoirs out of which its nitrogen might have originated can be classified by their isotopic fractionation. These reservoirs can basically be classified into four different areas, i.e.
\begin{description}
  \item[\textbf{Protosolar nebula:}] The solar wind and Jupiter resemble similar isotopic ratios of about $^{14}$N/$^{15}$N\,$\approx\,430-440$. These are the lightest known ratios in the solar system \citep{Marty2011,Owen2001} which are believed to represent primordial N$_2$ within the solar nebula.
  \item[\textbf{Cometary ices:}] Spectroscopic observations \citep{Rousselot2014,Shinnaka2014,Shinnaka2016} of several different comets from various dynamical groups in the outer solar system revealed isotopic ratios of NH$_3$ ices to be in the range of $^{14}$N/$^{15}$N\,$\approx 120-140$. These constitute the heaviest measured ratios in the solar system, except for localized ``hotspots'' in interplanetary dust particles (IDPs) and insoluble organic matter (IOM) that are extremely enriched in $^{15}$N compared to their surrounding \citep[e.g.][]{Busemann2006,Busemann2009,Briani2009,Bonal2010} with values as low as $^{14}$N/$^{15}$N\,$\approx$\,45 \citep{Briani2009}.
  \item[\textbf{N-bearing complex organics:}] IDPs \citep[e.g.][]{Joswiak2000,Busemann2009,Brownlee1995}, and IOM \citep[e.g.][]{Sandford2006,Alexander2007,Cody2011}, which can be found in comets, and chondrites, which are a class of primordial and primitive meteorites, contain complex refractory organic matter. These show a wide variety of isotopic ratios being in the range of $^{14}$N/$^{15}$N\,$\approx 160-320$ with an average value of $^{14}$N/$^{15}$N\,$ \approx 231$ \citep{Miller2019,McKeegan2006}.
\end{description}

Chondrites with their different subgroups (enstatite, ordinary, and carbonaceous chondrites) are further considered as one of the main building blocks of the terrestrial planets. With an isotopic ratio of $^{14}$N/$^{15}$N\,=\,$259\pm15$ \citep{Alexander2012}, this reservoir is the likely source of the Earth's atmospheric nitrogen \citep[e.g.][]{Marty2012}. Also Mars' interior and Venus' atmosphere show similar values \citep[e.g.,][]{Fueri2015}. However, organic matter is the main carrier of nitrogen within chondrites \citep[e.g.,][]{Alexander2007,Aleon2010}, and might therefore not be considered a distinct reservoir but an admixture of organic matter with another reservoir.

Since the Cassini era the atmospheric  $^{14}$N/$^{15}$N ratio at Titan is precisely known to be $167.7 \pm 0.6$ \citep{Niemann2010}, which lies between the reservoirs of ammonia ices and complex organics. To infer the initial building blocks of Titan, however, several different fractionation processes have to be taken into account and its effects have to be estimated through the whole evolution of Titan's atmosphere. The initial building blocks of Titan can therefore only be inferred if one studies these processes and retrieves the initial ratio of $^{14}$N/$^{15}$N.

There are two competing processes which fractionate nitrogen isotopes. Atmospheric escape preferentially removes the lighter isotope $^{14}$N from the atmosphere while the sequestration of nitrogen into nitriles preferentially removes the heavier isotope $^{15}$N \citep[see e.g.][]{Mandt2014,Krasnopolsky2016}. This ability to fractionate isotopes can be described with the so-called fractionation factor $f$; for $f>1$ the heavier isotope will preferentially be removed from the atmosphere, the lighter isotope for $f<1$. The change in isotopic fractionation from the present value to the initial value through $f$ further connects the initial inventory $n_0$, that is Titan's initial atmospheric inventory in this case, with the current inventory $n$, \textbf{Titan's} present-day atmospheric mass, through Rayleigh distillation \citep[e.g.][]{Lunine1999,Mandt2009,Mandt2014,Mandt2015}, i.e.
\begin{equation}\label{rayleigh}
  \frac{n_0}{n} = \left(\frac{R}{R_0}\right)^{\frac{1}{1-f}},
\end{equation}
where $R$ is the present-day $^{14}N$/$^{15}$N-ratio in Titan's atmosphere and $R_0$ the initial. If $R/R_0$ is less than one, then the present-day atmosphere is enriched in the heavy isotope, while the lighter isotope is \textbf{enriched }for $R/R_0>1$.

The fractionation factor $f$ has to be considered separately for different fractionation processes. In case of atmospheric escape, thermal (Jeans and hydrodynamic) and non-thermal (ion pickup, sputtering and suprathermal) escape processes fractionate differently. For Jeans escape the thermal fractionation factor $f_{\rm th}$ can be written as \citep{Mandt2009}
\begin{equation}\label{fracjeans}
  f_{\rm th} = \sqrt{\frac{m_1}{m_2}}\left[e^{(\lambda_1-\lambda2)}\frac{(1+\lambda_2)}{(1+\lambda_1)}\right],
\end{equation}
with $m_1$ and $m_2$ as the masses of the lighter and heavier \textbf{isotopologue}, and $\lambda_1$ and $\lambda_2$ as the respective Jeans escape parameters.
Even though escape through Jeans escape is highly mass-dependent, fractionation is rather slow due to low escape rates \citep[see e.g.][]{Mandt2014}). For small values of $\lambda$, Jeans escape turns into hydrodynamic escape and Equation \ref{fracjeans} converges towards
\begin{equation}\label{frachydro}
  f_{\rm th} \approx \sqrt{\frac{m_1}{m_2}}.
\end{equation}
Even though hydrodynamic escape can remove the whole bulk atmosphere, fractionation remains very low since the escape is less dependent on the mass of the particles.

Fractionation via non-thermal escape processes happens through diffusive separation of atmospheric particles from the homopause $r_{\rm h}$ to the exobase radius $r_{\rm exo}$. Due to lighter particles being more abundant at the exobase non-thermal escape will therefore -- just as thermal escape -- preferentially remove the lighter particle. This defines the non-thermal fractionation factor $f_{\rm nt}$ as follows \citep{Lunine1999,Mandt2009,Mandt2014}:
\begin{equation}\label{fracnont}
  f_{\rm nt} = exp\left[-\frac{g(\Delta z)(m_2-m_1)}{k_{\rm B}T}\right]
\end{equation}
with $g$ as the gravitational acceleration of the body and $\Delta z = r_{\rm exo}-r_{\rm h}$ as the distance between the homopause $r_{\rm h}$ and the exobase $r_{\rm exo}$.

For photochemical fractionation, which preferentially removes the heavier isotope from the atmosphere through sequestration into nitriles, the fractionation factor $f_{\rm ph}$ can be defined in two different ways \citep{Mandt2017},
\begin{enumerate}
\item through dividing the isotope ratio of the reactant ($R_{\rm re}$) by the isotope ratio of the product ($R_{\rm pr}$), i.e.
\begin{equation}\label{fracp1}
  f_{\rm ph1} = \frac{R_{\rm re}}{R_{\rm pr}},
\end{equation}
and
\item by determining the column rates for all the reactions involved in the production, and the loss of the isotopes, i.e.
\begin{equation}\label{fracp2}
  f_{\rm ph2} = \frac{F_2 \eta_1}{F_2 \eta_2},
\end{equation}
with $F_1$ and $F_2$ as total photochemical loss rates, and $\eta_1$ and $\eta_2$ as column densities of the lighter and heavier isotope.
\end{enumerate}

At Titan the main reaction product in which nitrogen will be sequestered is HCN. To approximate the fractionation of $^{14}$N/$^{15}$N through this process, by considering $f_{\rm ph1}$, it has to be assumed that all of the HCN originates from photodissociation of N$_2$. Therefore, to calculate $f_{\rm ph1}$, the $^{14}$N/$^{15}$N isotopic ratio of HCN has to be taken into account. This was measured by several different instruments \citep[][]{Marten1997,Marten2002,Gurwell2004,Vinatier2007,Courtin2011,Molter2016} with a mean value of $68.6\pm15.7$ \citep{Mandt2017}, resulting in $f_{\rm ph1}=(167.7\pm0.6)/(68.6\pm15.7)=2.44\pm0.56$. As \citet{Mandt2017} pointed out, however, all of the measurements of $^{14}$N/$^{15}$N in HCN were performed at solar minimum \citep[except for][]{Molter2016}. But during solar maximum conditions the photochemical loss rates of N$_2$ should be higher due to an increased solar irradiation, which will change the fractionation between N$_2$ and HCN. \citet{Mandt2017} therefore simulated $f_{rm ph1}$ values ranging from 2.05 for minimum, 1.67 for moderate and 1.59 for maximum conditions. For $f_{\rm ph2}$ they retrieved values from 1.86 for minimum to 1.56 for maximum conditions.

Since the photochemical fractionation factor is decreasing for increasing solar activity, the fractionation through photochemistry in the past might decrease as well due to the significant increase in solar EUV flux through time. Besides the potential availability of CH$_4$ in Titan's atmosphere this might hence be a second factor that diminishes the importance of photochemical fractionation as one gets farther back in time. We therefore chose $f_{\rm ph}=1.67=\mathrm{const.}$ as an average moderate value for calculating the total fractionation during the last 1000\,Myr.

If one wants to estimate the initial $^{14}$N/$^{15}$N fractionation of Titan's building blocks, all different fractionation factors, that is $f_{\rm th}$, $f_{\rm nt}$, and $f_{\rm ph}$, have to be combined and considered together through time as long as the respective factors are relevant processes acting in and on Titan's atmosphere. To calculate the total fractionation factor $f_{\rm tot}$ at any point in time, we followed \citet{Krasnopolsky2016}, i.e.
\begin{equation}\label{ftot}
f_{\rm tot} = \frac{\dot{M}_{\rm th}\times f_{\rm th} + \dot{M}_{\rm nt}\times f_{\rm nt} + \dot{M}_{\rm ph}\times f_{\rm ph}}{\dot{M}_{\rm th}+\dot{M}_{\rm nt}+\dot{M}_{\rm ph}}
\end{equation}
where $\dot{M}_{\rm th}$, $\dot{M}_{\rm nt}$, and $\dot{M}_{\rm ph}$ are the total atmospheric masses that are lost through thermal, non-thermal, and photochemical processes, respectively.

\begin{figure}
\begin{center}
\includegraphics[width=0.95\columnwidth]{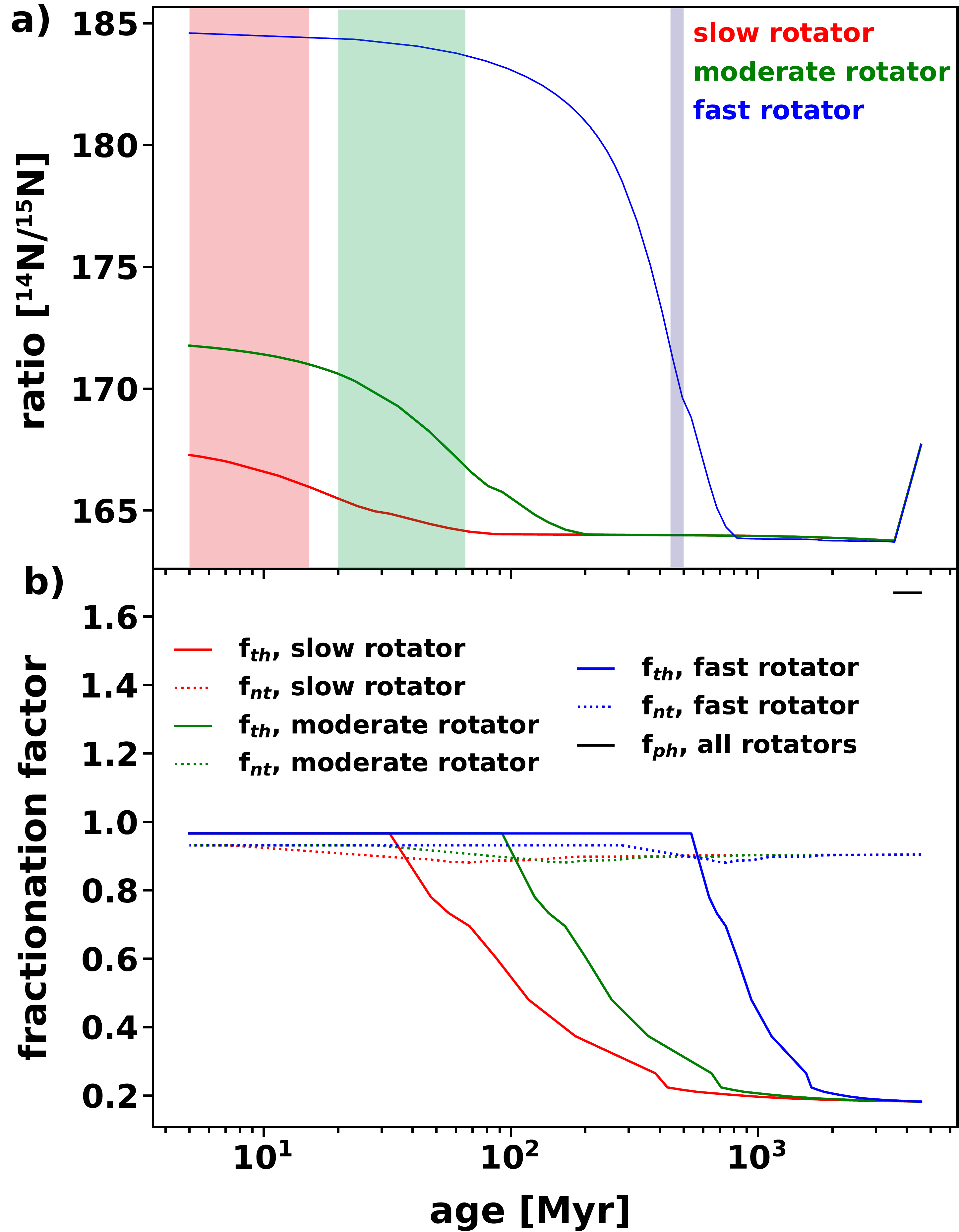}
\caption{a) Fractionation of $^{14}$N/$^{15}$N through time for a slow, moderate and fast rotator through thermal and non-thermal escape, as well as photochemical sequestration. The red, green and blue shaded areas represent the earliest possible times for the outgassing of Titan's nitrogen atmosphere for the different rotators if one assumes an endogenic origin of Titan's nitrogen from NH$_3$ ices and applies an adapted version of the model of \citet{Glein2015} with escape of nitrogen included (see Section~\ref{sec:out}). b) Evolution of the thermal ($f_{\rm th}$), non-thermal ($f_{\rm nt}$), and photochemical ($f_{\rm ph}$; only considered for the last 1000\,Myr) fractionation factor for a slow, moderate and fast rotator.}
\label{fig:frac}
\end{center}
\end{figure}

Fig.~\ref{fig:frac}a (for a description of the shaded areas see Section~\ref{sec:out}) shows the evolution of $\rm ^{14}N/^{15}N$ through time for a slowly, moderately and fast rotating Sun as calculated with Equation~\ref{ftot}. This plot can be roughly broken down into four distinct periods:
\begin{enumerate}
  \item During the last 1000\,Myr, the time in which we considered photochemical sequestration of N$_2$ into HCN to be relevant, the evolution is clearly dominated by fractionation through photolysis (with $f_{\rm ph}=1.67=\rm const.$), i.e. the $\rm ^{14}N/^{15}N$ ratio rises from $\approx\,164$ to its present-day value of 167.7. If we would consider this process to be relevant for a longer time, then the fractionation could have consequently been lower in the past. However, it is on the other hand unlikely that $f_{\rm ph}$ was constant through time; a decrease of $f_{\rm ph}$ due to an increase in EUV flux seems more likely \citep{Mandt2017}, which would consequently reduce the influence of this process in the past.
  \item A period in which $\rm ^{14}N/^{15}N$ changes insignificantly that starts with an EUV flux of about 30\,$\rm EUV_{\odot}$, i.e. after about 70\,Myr for a slowly, 170\,Myr for a moderately and 800\,Myr for a fast rotating Sun and ends with the onset of photochemical fractionation. Even though the thermal fractionation factor is very low during this period ($0.18 < f_{\rm th} < 0.37$; see Fig.~\ref{fig:frac}b) almost no fractionation occurs due to the close-to negligible corresponding thermal Jeans escape rates ($7 \times 10^{17}\,{\rm s^{-1}} < F_{\rm th} < 3 \times 10^{21}\,{\rm s^{-1}}$). For non-thermal escape, even though the escape rates are higher ($4 \times 10^{25}\,{\rm s^{-1}} < F_{\rm nt} < 6 \times 10^{25}\,{\rm s^{-1}}$), the corresponding fractionation factor $f_{\rm nt}$ stays close to $\approx 0.9$ over the whole duration of this period.
  \item All three rotator cases show a short transition phase from hydrodynamic to Jeans escape, in which thermal escape is capable to alter the initial $\rm ^{14}N/^{15}N$ ratio in Titan's atmosphere. While thermal escape increases, $f_{\rm th}$ converges towards one with a short period in which $\dot{M}_{\rm th}$ and $f_{\rm th}$ are both in a regime that will allow a significant alteration of the $\rm ^{14}N/^{15}N$ ratio. During this period $f_{\rm th}$ increases to 0.96, while the corresponding thermal escape increases to values above $10^{28}\,{\rm s^{-1}}$. Non-thermal escape plays a minor role in fractionating the isotopic ratio in Titan's atmosphere.
  \item The very early phase was shaped by hydrodynamic escape which fractionates $\rm ^{14}N/^{15}N$ insignificantly.
\end{enumerate}

If one assumes that Titan's atmosphere exists since the formation of the moon, then the initial fractionation of $^{14}$N/$^{15}$N would be $\approx 167$ for a slow, $\approx 172$ for a moderate and $\approx 185$ for a fast rotator, i.e. independently of the evolutional track of the young Sun one can assume that the $^{14}$N/$^{15}$N ratio should have been between 167\,--\,185. This value might be slightly different, if the role of photochemical sequestration was more or less pronounced over Titan's history. If we include photochemical fractionation over the whole time with $f_{\rm ph}=1.67=\mathrm{const.}$, we retrieve $^{14}$N/$^{15}$N\,$\geq$\,146, 151, and 172, for a slow, moderate, and a fast rotator, respectively. As described earlier, $f_{\rm ph}$ should decrease back in time due to the increasing EUV flux and CH$_4$ \textbf{may not have been} present over the whole duration. It seems, therefore, unlike that the initial value was that low, and these can be seen as absolute theoretical minima.

What does that mean for the initial building blocks of Titan's nitrogen atmosphere? In agreement with e.g. \citet{Mandt2014}, \citet{Krasnopolsky2016}, and \citet{Miller2019}, it seems unlikely that, in contrast to Earth, Titan's nitrogen originated from a chondritic reservoir since its isotope ratio \citep[$^{14}$N/$^{15}N=259\pm15$;][]{Alexander2012} is much \textbf{lighter }than the initial value of Titan's building blocks as retrieved from our simulations. Cometary ammonia ices on the other hand have a $^{14}$N/$^{15}N$ ratio of 120-140 \citep{Rousselot2014,Shinnaka2014,Shinnaka2016} which is slightly less than Titan's initial atmospheric value. It might, therefore, be reasonable to conclude that the main source of Titan's nitrogen could have likely been ammonia ices that were accreted by the moon, subsequently decomposed to N$_2$ in its interior and finally outgassed into its atmosphere, as suggested by \citet{Glein2015}. Ammonia ices as a primary source is also in agreement with several recent studies, such as \citet{Mandt2014} and \citet{Miller2019}. The slightly higher initial ratio of Titan's nitrogen could then either be explained by an admixture of its main source, ammonia ices, with one of the other reservoirs, such as N$_2$ accretion from the protosolar nebula \citep{Mandt2014} or accretion of N$_2$ bearing complex organics, as recently suggested by \citet{Miller2019}.

If, however, Titan's nitrogen mainly originated from ammonia ices that were subsequently decomposed and outgassed to build up its atmosphere, is it feasible that the atmosphere could have originated as early as the satellite itself? Or is it possible to infer some earliest potential outgassing age of the atmosphere? The whole nitrogen reservoir might have been lost if it was outgassed early, since there is only a restricted amount of NH$_3$ in Titan's interior \citep{Glein2015}. We will address this question in the following section.

\subsection{AN ENDOGENIC ORIGIN OF N$_2$ THROUGH DECOMPOSITION OF NH$_3$}\label{sec:out}
\begin{center}
\begin{table*}
\caption{Values for $f_{\rm{H_2O}}$ [decimal] for different interior models of Titan (after Fortes, 2012) and its related outgassing efficiency $\epsilon$ [\%].}
\begin{tabular}[t]{c c c c c c c } \hline
 & methane clathrate  &  pure water-ice  & light-ocean &  dense-ocean & iron sulphide inner core & pure iron inner core \\ \hline
$f_{\rm H_2O}$  & 0.211 & 0.303 & 0.25 & 0.355 & 0.296 & 0.298 \\
$\epsilon $ & 0.81 - 3.94 & 0.91 - 4.45 & 0.85 - 4.14 & 0.99 - 4.80 & 0.91 - 4.41 & 0.91 - 4.43 \\ \hline
\label{tabel1}
\end{tabular}
\end{table*}
\end{center}
\begin{center}
\begin{table}
\caption{Initial $^{14}$N/$^{15}$N ratio and total mass loss [$\rm M_{atm}$] for different solar rotational tracks, i. in case Titan's atmosphere originated directly at formation of the satellite, and ii. in case that the atmosphere was outgassed later-on from the deep interior through decomposition of NH$_3$ including the earliest possible time of outgassing [Myr].}
\begin{tabular}[t]{l|ccc} \hline
early origin: & & & \\
rotator & total loss & initial ratio  &  \\ \hline
slow   & 0.62   &   167.3 & \\
moderate & 2.22 & 171.9  & \\
fast   & 15.73 & 184.6  & \\ \hline
later origin: & & & \\
rotator & total loss & initial ratio & earliest outgassing \\ \hline
slow & 0.48\,-\,0.62 & 166.0\,-\,167.3 & 5\,-\,16  \\
moderate &  0.48\,-\,\textbf{2.02} & 167.2\,-\,170.6 & 20\,-\,67   \\
fast & 0.48\,-\,2.02 &   169.7\,-\,172.1 & 425 - 495  \\ \hline
\label{tabel2}
\end{tabular}
\begin{footnotesize}
fractionation factors: \\
$f_{\rm th}$ = 0.15 -- 0.94\\
$f_{\rm nt}$ = 0.88 -- 0.93\\
$f_{\rm ph}$ = 1.67\\
\end{footnotesize}
\end{table}
\end{center}
The hypothesis that Titan's initial atmosphere originated from ammonia ices deep within the moon was investigated by \citet{Glein2015} with a detailed analysis of chemical and isotopic data obtained from the Cassini-Huygens mission. Here it is assumed that the satellite's CH$_4$, N$_2$ and noble gases originated in a rocky core which is surrounded by an icy layer, with N$_2$ originating from the chemical decomposition of NH$_3$ through the reaction
\begin{equation}\label{eq:reaction}
\rm 2NH_3 \rightarrow N_2 + 3 H_2.
\end{equation}
Hydrothermal and cryovolcanic processes were then involved in the outgassing and build-up of Titan's atmosphere. \citet{Glein2015} performed mass balance and chemical equilibrium calculations and showed that a rocky core with a bulk noble gas content similar to that of CI carbonaceous chondrites would contain sufficient $^{36}$Ar and $^{22}$Ne to explain their observed abundances within Titan. Here we use the same approach as discussed in detail in \citet{Glein2015} and investigate how our simulated atmospheric escape processes constrain possible outgassing scenarios of Titan's N$_2$ atmosphere.

The outgassing efficiency $\epsilon$ determines the amount of N$_2$ that can be outgassed from the interior of Titan. It can be estimated through \citep{Glein2015}
\begin{equation}
\epsilon = {\tau \over \left( { \tau + \mu_{\mathrm{N_2}}\cdot
10^{ \left[ {\mathrm{log}_{10}
\left(
{ c_{\mathrm{Ar,rock}} \over y_{\rm{Ar,atm}} }
\times { (1-f_{\rm{H_2O}})  \over P_{\rm{atm}}  }
- {\tau \over \mu_{\rm{Ar}}
} \right) \pm 0.3 } \right] }} \right) } \: ,
\label{glein1}
\end{equation}
%
where $ \tau = ( 4 \pi / \rm{G} ) ( \rm{R}_{\rm{Titan}}^2 / \rm{M}_{\rm{Titan}} ) ^2 $, $\mu_{\rm{N_2}}$ and $\mu_{\rm{Ar}}$ are the molecular mass of N$_2$ and $^{36}$Ar, respectively, $ f_{\rm{H_2O}}$ is the mass ratio of water inside Titan, $P_{\rm{atm}}$ is the average atmospheric pressure (1.467 bar; Fulchignoni et al., 2005), $ c_{\rm{Ar,rock}} $ is the initial concentration of $^{36}$Ar in accreted rock, similar to CI chrondrites \citep[$(3.47 \pm 0.28 ) \times 10^{-8} \rm\, mol/kg$;][]{Pepin1991,Glein2015} and $ y_{\rm{Ar,atm}} $ is the observed mixing ratio \citep[$ (2.06 \pm 0.84) \times 10^{-7} $ for $^{36}$Ar;][]{Niemann2010,Glein2015}.

The saturation vapor pressures of argon and \textbf{hydrothermally} produced $\rm{N}_2 $ in form of clathrate hydrates and pure liquids under Titan's conditions are very similar \citep{Glein2015}. The small discrepancy between the two saturation vapor pressures is reflected by the uncertainty $\pm 0.3$ in Eq.~\ref{glein1} \citep{Glein2015}. For $f_{\rm H_2O}$ we took the values of the different structural models of Titan's interior as constructed by \citet{Fortes2012}. The various outgassing efficiencies $\epsilon$ for the different Fortes-models are described in Table \ref{tabel1}.

According to \citet{Glein2015}, the endogenic contribution of N$_2$ from NH$_3$ can be estimated from the following mass balance, i.e.
\begin{equation}
n_{\rm N_2,atm}^0=\epsilon \left( {\xi _{\rm{N_2}} \over 2 } \right) \left( {n_{\rm{NH}_3}^0 \over n_{\rm{H}_2O }^0} \right) \left( { f_{\rm{H_2O}} \over \mu_{\rm{H}_2\rm{O}} } \right) \mathrm{M}_{\rm{Titan}}
\label{N-Menge}
\label{outgassing1}
\end{equation}
with $ \xi_{\rm{N_2}} $ as the so-called reaction progress variable which specifies the yield of N$_2$ out of ammonia ices, and with ${n_{\rm{NH}_3}^0 }$ and ${n_{\rm{H}_2\rm{O}}^0 }$ as the entire initial amount of NH$_3$ and H$_2$O. Based on cometary observations \citep{Mumma2011,Glein2015} assumes that a reasonable ammonia to water ratio at Titan should be in the range of ${n_{\rm{NH}_3}^0 }/{n_{\rm{H}_2\rm{O}}^0 }=0.2\,-\,1.5\,\%$.

If one assumes that Titan's atmosphere originated from the decomposition of NH$_3$ ices in its interior then there are two competing processes that have to be taken into account, that is, outgassing and atmospheric escape. If the amount of nitrogen that escapes from the satellite over its history is higher than the amount that can be outgassed from its interior, Titan would have no atmosphere at present-day; if outgassing dominates, then the atmosphere can be sustained. The total inventory needed to build-up and sustain its present-day thick nitrogen atmosphere can, therefore, be assumed to be
\begin{equation}\label{inventory}
  n_{\rm N_2,atm}^0 \geq  n_{\rm N_2,atm} + n_{\rm N2,esc}
\end{equation}
with $n_{\rm N_2,atm}$ as the present-day amount of nitrogen in the atmosphere and $n_{\rm N2,esc}$ as the total amount of nitrogen that escaped from the satellite. Furthermore, an upper limit for the amount of nitrogen outgassed from the deep interior of Titan is given by $\xi_{\rm{N_2}}=1$ since the yield of N$_2$ out of NH$_3$ cannot be higher than 100\,\%. From Eq.~\ref{outgassing1} we can, thus, estimate the mass-balance parameter space that would be consistent with building-up the present-day nitrogen-dominated atmosphere around Titan through outgassing and sustaining it against atmospheric escape.

\begin{figure}
\begin{center}
\includegraphics[width=1.0\columnwidth]{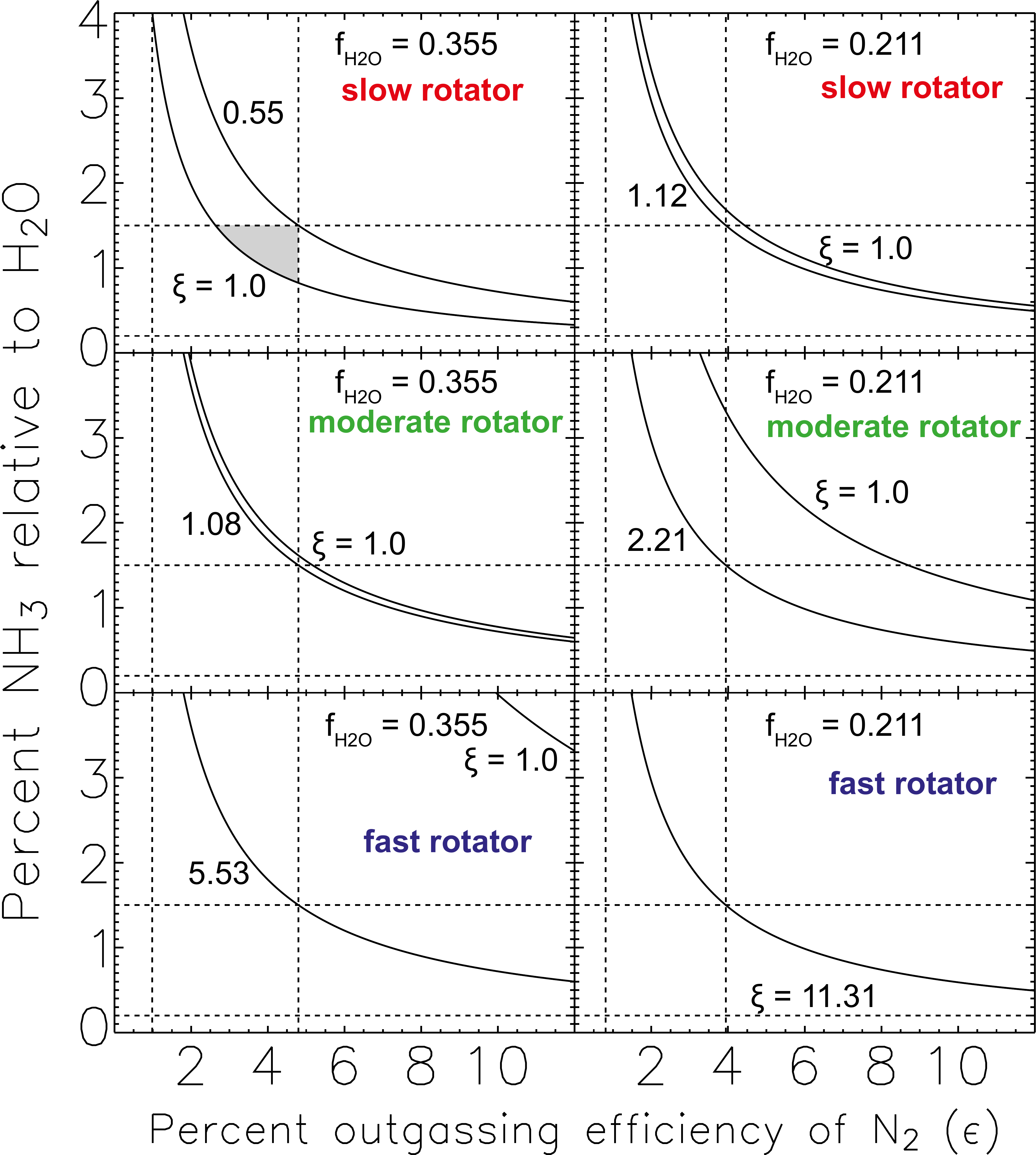}
\caption{The mass-balance parameter space for the outgassing of Titan's atmospheric N$_2$ through decomposition of accreted NH$_3$ in its deep interior according to Glein (2015) but with atmospheric escape for a slow (upper), moderate (middle), and fast rotator (lower panel) included and for the maximum($f_{\rm H_2O}=0.355$, left panels) and minimum ($f_{\rm H_2O}=0.211$, right) water ratios according to Fortes (2012). The plotted curves represent different values for the reaction progress variable $\xi$ which cannot be higher than 1 (i.e. 100\,\%). Therefore, the grey shaded area (only existing in the top left panel) represents the parameter space within which the existence of the present-day nitrogen atmosphere would be consistent. That means that if the atmosphere would have been outgassed right after the formation of Titan, it would have only survived until the present-day in case that the Sun was a slow rotator and with $f_{\rm H2O}=0.355$. In all other cases the escape would have been so strong that the available outgassed nitrogen reservoir could not have survived until present-day, meaning that the satellite would not have been able to sustain its thick envelope. In these cases Titan could only have an atmosphere if it was outgassed later-on or not all N$_2$ originated from ammonia ices.}
\label{fig:outgassing}
\end{center}
\end{figure}

Fig.~\ref{fig:outgassing} shows the parameter space for building-up and sustaining Titan's present-day atmosphere for the maximum ($f_{\rm H_2O}\,=\,0.355$, left panels) and minimum ($f_{\rm H_2O}\,=\,0.211$, right) water to total mass ratios according to \citet{Fortes2012} and for a slow (upper), moderate (middle), and fast rotator (lower panel). The  dashed lines show reasonable limits for the NH$_3$/$\rm H_2O$ ratio and the outgassing efficiency $\epsilon$. The grey-shaded area in the upper-left panel illustrates the parameter space within which Titan's present-day atmosphere would be consistent in case that the atmosphere would have been outgassed right after the formation of the satellite and with atmospheric escape included. As can be clearly seen, only for a very high water ratio $f_{\rm H_2O}=0.355$ and for a slow rotator the present-day atmosphere could have been built-up and would not have been lost into space. For any other cases, Titan's atmosphere would not have been able to survive, if it would have been outgassed directly after formation of the satellite. One can, therefore, conclude that, if its N$_2$ originated from ammonia ices, then the Sun was either a slow rotator and/or the atmosphere was outgassed later-on.

Table~\ref{tabel2} shows the total mass loss over time through escape and the initial $^{14}$N/$^{15}$N ratio if Titan's nitrogen atmosphere either originated at the formation of the satellite or later-on through outgassing from the deep interior. In the latter case Table~\ref{tabel2} also shows the earliest possible time of outgassing for the different rotational evolution tracks of the young Sun. If the assumption that Titan's atmosphere originated mainly endogenically through decomposition of NH$_3$ into N$_2$, then the atmosphere could not have been outgassed earlier than 5~Myr for a slowly, 20~Myr for a moderately, and 425~Myr after formation of the solar system for a fast rotating young Sun; it has to be noted, however, that this result does not give a boundary for the latest outgassing of the atmosphere. In such a case the maximum amount of nitrogen lost from Titan's atmosphere would be 0.62\,M$_{\rm atm}$ for a slow and 2.02\,M$_{\rm atm}$ for a moderate and fast rotator. The initial fractionation of $^{14}$N/$^{15}$N of Titan's building blocks would then accordingly be less than 167.3, 170.6, and 172.1, respectively. The corresponding earliest times for outgassing are also illustrated as red, green, and blue boxes in Fig.~\ref{fig:frac}a.

An initial fractionation of Titan's atmospheric $^{14}$N/$^{15}$N ratio below 172 indeed most closely relates to ammonia ices being the main reservoir for its nitrogen, which is in agreement with other recent studies such as \citet{Mandt2014}, \citet{Krasnopolsky2016}, and \citet{Miller2019}. As described in Section~\ref{sec:frac}, spectral observations of NH$_3$ within different comets \citet{Rousselot2014,Shinnaka2014,Shinnaka2016} show a fractionation of $^{14}$N/$^{15}$N$\,\approx\,120-130$. This is close, but most likely below the initial fractionation of Titan's N$_2$. Therefore, it seems likely that ammonia ices weren't the only building blocks that were involved in the accumulation of nitrogen at Titan. Some fraction of its N$_2$ could have originated from the solar nebula, chondrites, or nitrogen bearing refractory organics. If Titan accumulated N$_2$ from the solar nebula, however, it could have only been a tiny fraction which would not have survived the strong atmospheric escape early-on in the evolution of the solar system. The elevated fractionation compared to NH$_3$ at Titan could, hence, only originate from another reservoir. Alternatively, the role of photochemical sequestration at Titan could have been more important in the past than assumed in our simulation, which could also decrease the $^{14}$N/$^{15}$N ratio even further. For decreasing it to $\approx\,120-130$, the value of ammonia ices, photochemical sequestration must have worked constantly over the whole lifetime of the solar system \citep{Krasnopolsky2016}. This scenario, however, seems very unlikely due to the arguments outlined above -- a lower $f_{\rm ph}$ for higher EUV fluxes and the continuous availability of CH$_4$ in Titan's atmosphere.

\citet{Miller2019} estimated the mean $^{14}$N/$^{15}$N ratio of refractory organics and compiled ratios from different nitrogen bearing sources of the outer solar system that are believed to be carriers of refractory organic nitrogen. These are i) IDPs, which are believed to originate from comets and to represent primitive solar system material, ii) Stardust samples from comet Wild 2, and iii) IOM) from carbonaceous chondrites that likely formed in the outer solar system such as CR chondrites and the CH/CB Isheyevo chondrite. These nitrogen bearing sources have $^{14}$N/$^{15}$N ratios ranging from 160 to 320 with an average of $^{14}$N/$^{15}$N\,=\,231 \citep{Miller2019,McKeegan2006}. According to \citet{Miller2019}, N$_2$ from refractory organics seems to be an additional likely source of nitrogen at Titan besides NH$_3$ ices. In particular because a mixture of approximately 50/50 between those two resources would also explain the $^{36}$Ar/$^{14}$N ratio in Titan's atmosphere -- an atmosphere purely originating from cometary NH$_3$ ices would have an $^{36}$Ar/$^{14}$N ratio that would be higher, whereas an atmosphere purely originating from refractory organics would have a ratio that would be lower than Titan's. Furthermore, \citet{Miller2019} estimated that the amount of nitrogen within refractory organics in Titan's interior would amount to up to $\approx$\,133 times the present-day atmospheric mass of N$_2$.

Our simulations are in agreement with a mixture of these two different reservoirs, -- ammonia ices and N bearing refractory organics -- with both mainly originating from accreted comets. Depending on the on-turn of the nitrogen degassing from the interior into the atmosphere and the therewith connected total escape of nitrogen, however, the mixture in our model might be relatively different. While $^{14}$N would be lost from the atmosphere, $^{36}$Ar, which is significantly heavier than N$_2$ or even N, would accumulate in Titan's atmosphere, which in turn would elevate the atmospheric $^{36}$Ar/$^{14}$N ratio from its pristine value. Even though we did not simulate the escape of $^{36}$Ar, it is arguably likely that the escape of nitrogen can at least partially account for the observed $^{36}$Ar/$^{14}$N ratio; the admixture between nitrogen originating from ammonia ices and from refractory organics might, thus, not be $\approx$\, 50/50 as suggested by \citet{Miller2019} but dominated by the NH$_3$ reservoir. A simulation of the escape of the noble gases from Titan's atmosphere through its history will be needed in the future, however, to answer this question more accurately.

In addition, future isotopic measurements will be crucial to further constrain the origin and evolution of Titan's atmosphere. These include the measurement of $^{14}$NH$_3$/$^{15}$NH$_3$ at Titan's surface ices, as already suggested by \citet{Glein2015}, but also the precise measurement of the noble gas abundances and ratios in its atmosphere. The isotopic ratios of $^{36}$Ar/$^{38}$Ar and $^{20}$Ne/$^{22}$Ne, for instance, can tell us something about different fractionation processes over the history of Titan's atmosphere and when it might have originated. If it originated early-on, at a time when thermal escape was yet significant, the escaping nitrogen could have dragged Ar and Ne isotopes from its atmosphere, thereby fractionating the more easily dragged lighter isotopes from the heavier one. This process likely played a crucial role in the early history of Venus and Earth when their potential primordial hydrogen-dominated atmospheres escaped, dragged away and fractionated their atmospheric noble gases \citep[e.g.,][]{Lammer2020}, but this process could have also significantly fractionated $^{36}$Ar/$^{38}$Ar, $^{20}$Ne/$^{22}$Ne, and particularly $^{22}$Ne/$^{36}$Ar in Titan's atmosphere, if it was already existing at the time of strong thermal escape. Sputtering could further have specifically fractionated the lighter Ne isotopes at a later stage in Titan's history. Determining these different isotopic ratios via simulations and with a future space mission, such as Dragonfly with the Dragonfly Mass Spectrometer DraMS \citep{Lorenz2018}, can, therefore, tell us a lot about the evolution of Titan's atmosphere. If at some point in \textbf{the} future the \textbf{currently} unknown noble gas ratios in comets could also be determined, these, together with a thorough modeling of the diverse fractionation processes in Titan's atmosphere would shed additional light on the origin of the noble gases of Saturn's biggest satellite.

\section{CONCLUSION}

We applied a 1D upper atmosphere EUV radiation absorption and hydrodynamic escape model to study thermal escape of nitrogen from Titan's atmosphere over time. Our model retrieved very low escape rates for the present-day but significant escape very early-on in the history of the solar system. For 100 and 400\,EUV$_{\rm \odot}$ the thermal escape rates of nitrogen would be as high as $1.4\times10^{28}\rm s^{-1}$ and $4.6\times10^{29}\rm s^{-1}$, respectively, whereas today the thermal escape rate would only be $7.5\times10^{17}\rm s^{-1}$. Depending on whether the Sun was a slow, moderate or fast rotator, thermal escape would have been the dominant escape process for the first 100\,Myr up to 1000\,Myr, in case that Titan's atmosphere originated together with the formation of the satellite. In such a case the total loss of nitrogen would have ranged from 0.6 times the present-day atmospheric mass for a slow to 15.7 times the present-day atmospheric mass for a fast rotator. However, it again has to be noted that the Sun most likely should have been a slowly, or at maximum, between a slowly and moderately rotating young G-type star which favors the lower values for the cumulative loss. We also estimated the loss of nitrogen through non-thermal escape processes and found them to be negligible compared to the loss through thermal escape. If one considers an endogenic origin of Titan's N$_2$ from NH$_3$ ices that were decomposed in its deep interior, then its atmosphere could not have originated directly after the formation of the satellite except if the Sun indeed was a slow rotator and the water to rock ratio at the satellite is considerably high ($f_{\rm H_2O}\,\approx\,0.355$). In all other cases, it should have originated later. The thermal escape would have been so strong that the atmosphere could not have been sustained until the present-day which would even be pronounced if one also includes atmospheric escape via impact erosion. An endogenic origin of Titan's N$_2$ at least partially through ammonia ices, is also consistent with the estimated initial fractionation of its atmospheric N$_2$, which should have been $^{14}$N/$^{15}$N$\,\leq\,166\,--\,172$. Since this ratio, however, is slightly above the ratio of cometary ammonia ices, it seems also likely that part of Titan's nitrogen, likely below 50\% originated from refractory organics. Such a combination together with atmospheric escape could also explain the $^{36}$Ar/$^{14}$N ratio in Titan's atmosphere. Future isotopic measurements such as $^{14}$NH$_3$/$^{15}$NH$_3$ at Titan's surface ices and a precise measurement of the noble gas abundances and ratios in its atmosphere combined with simulations on their escape will be crucially important to gain further insides into the origin and evolution of Titan's thick nitrogen-dominated atmosphere.

\section*{APPENDIX}

For integrating the system of equations on time, we apply a two-step MacCormack numerical scheme.
Finite difference equations for the total density, ions and atomic nitrogen densities can be written in the following vector form:
 \begin{eqnarray}
  \hat{n}_{i}^{k+1/2} =  \hat{n}_{k i}^k - \frac{\Delta t}{r_i^2}\frac{( \hat{\Gamma}_{i+1}^k  -
 \hat{\Gamma}_{i}^k  )}{(r_{i+1}-r_i)} + \hat{S}_{i}^k, \\
  \hat{n}_i^{k+1} = \frac{1}{2} (\hat {n}_i^{k+1/2}+\hat{n}_i^{k}) - \nonumber \\
 0.5\frac{\Delta t}{r_i^2}\frac{( \hat{\Gamma}_{i}^{k+1/2}  -
 \hat{\Gamma}_{i-1}^{k+1/2} )}{(r_i -r_{i-1})} + \hat{S}_{i}^{k+1/2} ,
\end{eqnarray}
where $\hat {n}$ = $ (\rho, n_{\rm N}, n_{\rm N^+}, n_{\rm N_2^+})$,  $\hat \Gamma = \hat{n} V r^2$, $\hat{S} = (0, S_{\rm N}, S_{\rm N^+}, S_{\rm N_2^+})$ .

The momentum finite difference equations can be written as
\begin{eqnarray}
{(\rho V)}_i^{k+1/2} = (\rho V)_i^k -
\frac{\Delta t}{r_i^2}\frac{( \Pi_{i+1}^k  -
 \Pi_i^k  )}{(r_{i+1}-r_i)} -  \nonumber \\
 0.5\frac{\Delta t}{r_i^2}(r_i^2\rho_i^k +r_{i+1}^2\rho_{i+1}^k)\frac{(U_{i+1}-U_i)}{(r_{i+1}-r_i)}  -   \nonumber \\
 0.5\frac{\Delta t}{r_i^2}{(r_i^2+r_{i+1}^2)} \frac{(P_{i+1}^k -P_{i}^k)}{(r_{i+1}-r_i)}  , \\
 {(\rho V)}_i^{k+1} = 0.5  [(\rho V)_i^{k+1/2} + (\rho V)_i^{k}] -\nonumber \\
  0.5\frac{\Delta t}{r_i^2}\frac{( \Pi_{i}^{k+1/2}  -
 \Pi_{i-1}^{k+1/2} )}{(r_i-r_{i-1})} +  \\
 -0.25\frac{\Delta t}{r_i^2}(r_i^2\rho_i^{k+1/2}+r_{i-1}^2\rho_{i-1}^{k+1/2}) \frac{(U_{i}-U_{i-1})}{(r_i - r_{i-1})}  -  \nonumber \\
 0.25\frac{\Delta t}{r_i^2}(r_i^2+r_{i-1}^2)\frac{( P_i^{k+1/2}-P_{i-1}^{k+1/2})}{(r_i - r_{i-1})}   ,
 \end{eqnarray}
where $\Pi = \rho V^2 r^2 $ .

And the energy conservation equation is approximated as follows:
\begin{eqnarray}
{W}_i^{k+1/2} = W_i^k - {\Delta t}\frac{( G_{i+1}^k  -
 G_i^k)}{(r_{i+1}-r_i)} +  {\Delta t} r_i^2 Q_{\rm{EUV} i}^k   , \\
 {W}_i^{k+1} = 0.5(W_i^{k+1/2} + W_i^{k}) - \nonumber \\
 0.5{\Delta t}\frac{( G_{i}^{k+1/2}  -  G_{i-1}^{k+1/2})}{(r_i-r_{i-1})}
  0.5{\Delta t} r_i^2 Q_{\rm{EUV} i}^{k+1/2} +  \nonumber\\
  \frac{\Delta t}{(r_{i+1}-r_{i-1}) }  \left[ r_{i+1/2}^2  \chi_{i+1/2} \frac{(T_{i+1}^{k+1}-T_{i}^{k+1})}{(r_{i+1}-r_i)} \right. \nonumber\\
  \left. - r_{i-1/2}^2 \chi_{i-1/2} \frac{(T_{i}^{k+1}-T_{i-1}^{k+1})}{(r_{i}-r_{i-1})} \right ] ,
\end{eqnarray}
where $ W = r^2 (\rho V^2 /2 + \rho U + E )$, $G = V (W + P r^2)$.

\section*{Data Availability}

\textbf{The data underlying this article will be shared on reasonable request to the corresponding author.}

\section*{ACKNOWLEDGMENTS}
The authors acknowledge the support by the FWF NFN project S11601-N16 `Pathways to Habitability: From Disks to Active Stars, Planets and Life', and the related FWF NFN subprojects, S11604-N16 'Radiation \& Wind Evolution from T Tauri Phase to ZAMS and Beyond', and S11607-N16 `Particle/Radiative Interactions with Upper Atmospheres of Planetary Bodies Under Extreme Stellar Conditions'. H. Lammer, P. Odert and N. V. Erkaev acknowledges also support from the FWF project P25256-N27 `Characterizing Stellar and Exoplanetary Environments via Modeling of Lyman-$\alpha$ Transit Observations of Hot Jupiters'. NVE and VAI acknowledge support by the Russian Science Foundation project No 18-12-00080. We finally thank two anonymous referees who helped to improve the manuscript a lot.

\bibliography{Erkaev-et-al_Titan_arxiv}

\end{document}